%% file: final.tex
\documentclass[acmtog,nonacm,authorversion,timestamp]{acmart}
\pdfoutput=1
\PassOptionsToPackage{hyphens}{url}

\newcommand{\finalversion}{1}

\makeatletter                   
\def\mdseries@tt{m}             
\makeatother                    

\usepackage{breakurl}           

\input{headers/01-metaacm}

\input{headers/03-graphics}
\input{headers/10-package}
\input{headers/13-usenatbib}
\input{headers/20-macros}
\input{headers/21-symbols}

\input{headers/22-acronyms}

\begin{document}

\input{body/00-title}
\input{body/01-authors}
\input{body/02-abstract}

\input{body/03-classification}
\input{body/04-teaser}

\maketitle

\input{body/10-introduction}

\input{body/20-related-work}
\input{body/30-method}
\input{body/35-implementation}

\input{body/40-results}

\input{body/90-conclusion}

{
	\input{headers/99-tweakbib}
	\bibliographystyle{acmart}
	\bibliography{biblio/biblio} 
}


\end{document}

%% file: headers/01-metaacm.tex
\setcitestyle{square}
\setcitestyle{nosort}

\let\oldparagraph\paragraph
\renewcommand{\paragraph}[1]{\oldparagraph{\textbf{#1}}}

\let\oldsubparagraph\subparagraph
\renewcommand{\subparagraph}[1]{\oldsubparagraph{\emph{#1}}}

\acmJournal{TOG}
\acmYear{}
\acmVolume{}
\acmNumber{}
\acmArticle{}
\acmMonth{7} 
\acmDOI{}
\acmSubmissionID{} 

\setcopyright{none}

%% file: headers/03-graphics.tex
\graphicspath{
	{figs/raster/}
	{figs/handdrawn/}
	{figs/vector/}
	{figs/photos/}
        {figs/elements/}
        {figs/render/}
	{../../Printex/doc/figs/raster/}
	{../../Printex/doc/figs/handdrawn/}
}

%% file: headers/10-package.tex
\usepackage[l2tabu,orthodox]{nag}

\usepackage[T1]{fontenc}
\usepackage[utf8]{inputenc}
\usepackage[shrink=10]{microtype}

\usepackage{amsthm}
\usepackage{amssymb}
\usepackage{mathtools} 
\usepackage{textcomp}
\usepackage{siunitx} 
\usepackage{physics} 
\usepackage{bm}
\usepackage{stmaryrd}

\usepackage{ifthen}
\usepackage{hyperref}
\usepackage{cleveref}
\usepackage{csquotes}
\usepackage{calc}
\usepackage[usenames,dvipsnames]{xcolor}

\usepackage{alltt}
\usepackage[ruled,vlined,onelanguage,linesnumbered]{algorithm2e}

\usepackage{array}
\usepackage{booktabs}
\usepackage{multirow}

\usepackage[frozencache]{minted}

\usepackage{float}
\usepackage{wrapfig} 
\usepackage{graphicx}
\usepackage[caption=false]{subfig} 
\usepackage{tikz}
\usetikzlibrary{decorations.pathmorphing,patterns,shadows,calc}

\usepackage{etoolbox}

%% file: headers/13-usenatbib.tex
\newcommand{\textcite}[1]{\citet{#1}}

%% file: headers/20-macros.tex
\definecolor{AudioColor}{rgb}{0.56,0.34,0.62}
\definecolor{DeadlineColor}{rgb}{0.9,0.4,0}

\newcommand{\todo}[1]{\textcolor{red}{TODO: #1 $\qed$}}

\newcommand{\warning}[1]{{\itshape\color{red} #1}}
\newcommand{\note}[1]{{\itshape\color{blue} #1}}
\newcommand{\deadline}[1]{{\bf\color{DeadlineColor} ETA: #1}}
\newcommand{\noref}[1]{{\leavevmode\itshape\color{red}[ref:#1]}}

\providecommand{\finalversion}{1}
\ifthenelse{\equal{\finalversion}{1}}
{
	\renewcommand{\todo}[1]{}
	\renewcommand{\warning}[1]{}
	\renewcommand{\note}[1]{}
	\renewcommand{\deadline}[1]{}
	\renewcommand{\noref}[1]{}
}
{}

\newcolumntype{M}[1]{>{\centering\arraybackslash}m{#1}}

\definecolor{forestgreen}{rgb}{0.13,0.54,0.13}

\SetCommentSty{algocomment}
\SetKw{Break}{break}
\SetKw{Continue}{continue}
\SetKw{Or}{or}

\crefname{algocf}{alg.}{algs.}
\Crefname{algocf}{Algorithm}{Algorithms}

\definecolor{darkblue}{rgb}{0,0,.5}
\hypersetup{
	unicode=true,
	colorlinks=true,
	citecolor=forestgreen, 
	linkcolor=darkblue, 
	urlcolor=darkblue, 
    breaklinks=true,            
}

\newcommand{\filename}[1]{\url{#1}}
\newcommand{\foldername}[1]{\url{#1}}

\newcommand{\sharedVideosUrl}{https://drive.google.com/drive/folders/0B_DXj6bIWKPvNkJwbFM3WnJUY3M?usp=sharing}

\newcommand{\sharedMeshesUrl}{https://drive.google.com/open?id=0B_DXj6bIWKPvVjFQZ2VvaVp5eW8}

\ifdefined\siggraph
\LetLtxMacro\oldcaption\caption
\renewcommand{\caption}[2][]{\oldcaption[#1]{{\fontsize{\captionfontsize}{\captionfontskip}\selectfont #1} {\fontsize{\captionfontsize}{\captionfontskip}\selectfont #2}}}
\newcommand{\Caption}[2]{\caption[#1]{#2}}
\else
\SetAlFnt{\small}
\SetAlCapFnt{\small}
\SetAlCapNameFnt{\small}
\SetAlCapHSkip{0pt}
\IncMargin{-\parindent}
\newcommand{\Caption}[2]{\caption[#1]{{\em #1} #2}}
\fi

\setminted{tabsize=4, autogobble=true, mathescape, breaklines=true, fontsize=\footnotesize}

\crefname{lstlisting}{listing}{program}
\Crefname{lstlisting}{Listing}{Program}

\newcommand{\iflabelexists}[3]{\ifcsundef{r@#1}{#3}{#2}}

%% file: headers/21-symbols.tex
\let\originalleft\left
\let\originalright\right
\renewcommand{\left}{\mathopen{}\mathclose\bgroup\originalleft}
\renewcommand{\right}{\aftergroup\egroup\originalright}

\renewcommand{\geq}{\geqslant}
\renewcommand{\leq}{\leqslant}

\newcommand{\eqdef}{\stackrel{\text{\tiny def}}{=}}

\newcommand{\N}{\mathcal{N}}

\newcommand{\reals}{\mathbb{R}}

\DeclareMathOperator*{\minimize}{minimize}
\DeclarePairedDelimiter\size{\lvert}{\rvert}

\newcommand{\vect}[1]{\mathbf{#1}}
\newcommand{\mat}[1]{\mathbf{#1}}
\newcommand{\identity}{\mat{I}}

\newcommand{\transposeSymbol}{\mathstrut\scriptscriptstyle{\bm{\top}}}
\newcommand{\transpose}[1]{#1^{\transposeSymbol}}

\newcommand{\gridCell}{i}
\newcommand{\allGridCells}{{\mathsf G}}
\newcommand{\gridCellDomain}{\Omega_\gridCell}
\newcommand{\compliance}{\mathcal{C}}
\newcommand{\complianceFunc}[1]{\compliance\left(#1\right)}
\newcommand{\densityGridCellSymbol}{x}
\newcommand{\densityGridCell}{\densityGridCellSymbol_\gridCell}
\newcommand{\allDensities}{\vect{\densityGridCellSymbol}}
\newcommand{\volumeGridCell}{\mathrm{Vol}(\gridCell)}

\newcommand{\externalForces}{\mathbf{f}_{\mathrm{ext}}}
\newcommand{\displacements}{\mathbf{u}}
\newcommand{\localDisplacements}{\mathbf{u}_\gridCell}
\newcommand{\stiffnessMatrix}{\mathbf{K}}
\newcommand{\stiffnessMatrixBase}{\stiffnessMatrix_0}

\newcommand{\lieAlgebra}{so} 
\newcommand{\rotationGroup}{SO}

\newcommand{\rotationMatrix}{R}

\newcommand{\element}{e}

\newcommand{\perElement}[1]{#1^{\element}}

\newcommand{\dofTranslation}{t}
\newcommand{\dofsTranslation}{\vect{\dofTranslation}}
\newcommand{\dofRotation}{\gamma}
\newcommand{\dofsRotation}{\vect{\dofRotation}}

\newcommand{\elementTranslation}{\perElement{\mat{T}}}
\newcommand{\elementRotation}{\perElement{\mat{R}}}
\newcommand{\elementLinear}{\perElement{\mat{A}}}

\newcommand{\dofSamplePos}{y}
\newcommand{\dofsSamplePos}{\vect{\dofSamplePos}}

\newcommand{\allDofsSamplePos}{\mat{Y}}
\newcommand{\allDofsSamplePosElement}{\perElement{\allDofsSamplePos}}

\newcommand{\dofAngularPos}{\omega}
\newcommand{\dofsAngularPos}{\bm{\omega}}
\newcommand{\deltaSamplePos}{\bm{\delta}} 
\newcommand{\rootedTreeElement}{\perElement{\mathcal T}}

\newcommand{\dofsSymbol}{\Theta}
\newcommand{\allDofs}{\mat{\dofsSymbol}}
\newcommand{\allDofsElement}{\perElement{\allDofs}}
\newcommand{\oneDof}[1]{\theta_{#1}}

\newcommand{\sample}{s}
\newcommand{\otherSample}{{s'}}
\newcommand{\allSamples}{\mathcal S}

\newcommand{\positionSymbol}{p}
\newcommand{\currentPosition}{\vect{\positionSymbol}}
\newcommand{\samplePosition}{\vect{\positionSymbol}_\sample}
\newcommand{\samplePositionCoord}[1]{\positionSymbol_{\sample_{#1}}}

\newcommand{\allSamplePos}{\mat{P}}
\newcommand{\allSamplePosElement}{\perElement{\allSamplePos}}
\newcommand{\sampleRadiusSymbol}{r}
\newcommand{\sampleRadius}{\sampleRadiusSymbol_\sample}

\newcommand{\parent}[1]{\mathrm{pred}[#1]}

\newcommand{\materialDensitySymbol}{\rho}
\newcommand{\densityAllSamples}{\materialDensitySymbol}
\newcommand{\densitySampleFunc}{\materialDensitySymbol_\sample}
\newcommand{\densityElementFunc}{\materialDensitySymbol_\element}
\newcommand{\densityOtherSampleFunc}{\materialDensitySymbol_{\otherSample}}
\newcommand{\rbfWidth}{\alpha}
\newcommand{\rbfSlope}{\beta}
\newcommand{\gaussPoint}{\widetilde{\vect{\positionSymbol}}}
\newcommand{\gaussWeight}{\omega}

\newcommand{\dofsToSamplePos}{h}
\newcommand{\samplePosToDensities}{g}
\newcommand{\densitiesToScalar}{f}

\newcommand{\jacobianMatrix}[1]{\mat{J}_{#1}}

\newcommand{\graphEdges}{E}
\newcommand{\graphWeight}{W}

\newcommand{\distanceToBoundary}[1]{d_{\delta \Omega}(#1)}

\newcommand{\insideElement}[1]{\mathbf{1}_{\element}(#1)} 

\newcommand{\domainFunction}{f_{\mathrm{domain}}}



%% file: headers/22-acronyms.tex
\newcommand{\cLang}{C}

\newcommand{\libigl}{\texttt{libigl}}



%% file: body/00-title.tex
\title{Printable Aggregate Elements}

%% file: body/01-authors.tex
\newcommand{\unilo}{Universit\'{e} de Lorraine}
\newcommand{\inria}{INRIA}
\newcommand{\nyu}{NYU}
\newcommand{\hku}{University of Hong Kong}
\newcommand{\ntop}{nTopology}
\newcommand{\adobe}{Adobe Research}


\author{J\'{e}r\'{e}mie Dumas}
\orcid{0000-0001-7304-9882}
\affiliation{\institution{\inria{}}}
\affiliation{\institution{\hku{}}}
\affiliation{\institution{\ntop{}}}

\author{Jon\`{a}s Mart\'{i}nez}
\affiliation{\institution{\inria{}}}
\affiliation{\institution{\unilo{}}}

\author{Sylvain Lefebvre}
\affiliation{\institution{\inria{}}}
\affiliation{\institution{\unilo{}}}

\author{Li-Yi Wei}
\affiliation{\institution{\hku{}}}
\affiliation{\institution{\adobe{}}}

\renewcommand{\shortauthors}{Dumas, Mart\'{i}nez, Lefebvre, and Wei}


%% file: body/02-abstract.tex
\begin{abstract}
	Aggregating base elements into rigid objects such as furniture or sculptures is a great way for designers to convey a specific look and feel. Unfortunately, there is  no existing solution to help model structurally sound aggregates.
	The challenges stem from the fact that the final shape and its structural properties emerge from the arrangements of the elements, whose sizes are large so that they remain easily identifiable.
	Therefore there is a very tight coupling between the object shape, structural properties, and the precise layout of the elements.

	We present the first method to create aggregates of elements that are structurally sound and can be manufactured on 3D printers.
	Rather than having to assemble an aggregate shape by painstakingly positioning elements one by one, users of our method only have to describe the structural purpose of the desired object. This is done by specifying a set of external forces and attachment points. The algorithm then automatically optimizes a layout of user-provided elements that answers the specified scenario.
	The elements can have arbitrary shapes: convex, concave, elongated, and can be allowed to deform.

	Our approach creates connections between elements through small overlaps preserving their appearance, while optimizing for the global rigidity of the  resulting aggregate. We formulate a topology optimization problem whose design variables are the positions and orientations of individual elements. Global rigidity is maximized through a dedicated gradient descent scheme. Due to the challenging setting -- number of elements, arbitrary shapes, orientation, and constraints in 3D -- we propose several novel steps to achieve convergence.
\end{abstract}

%% file: body/03-classification.tex
\ccsdesc[500]{}

\keywords{Additive manufacturing, topology optimization, discrete elements.}

\thanks{
	This work is supported by ERC grant ShapeForge (StG-2012-307877).
}

%% file: body/04-teaser.tex

  \begin{teaserfigure}
  \centering
  \subfloat[elements]{
        \label{fig:teaser:exemplar}
        \begin{minipage}[c][3.5cm]{0.12\linewidth}
          \centering
          \vspace{2ex}
    \includegraphics[width=0.30\linewidth]{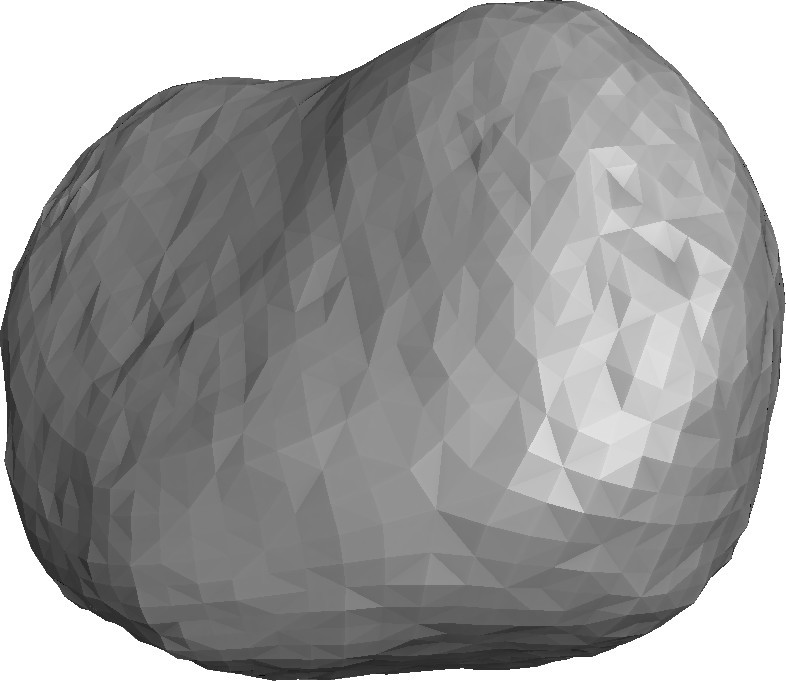}
    \includegraphics[width=0.30\linewidth]{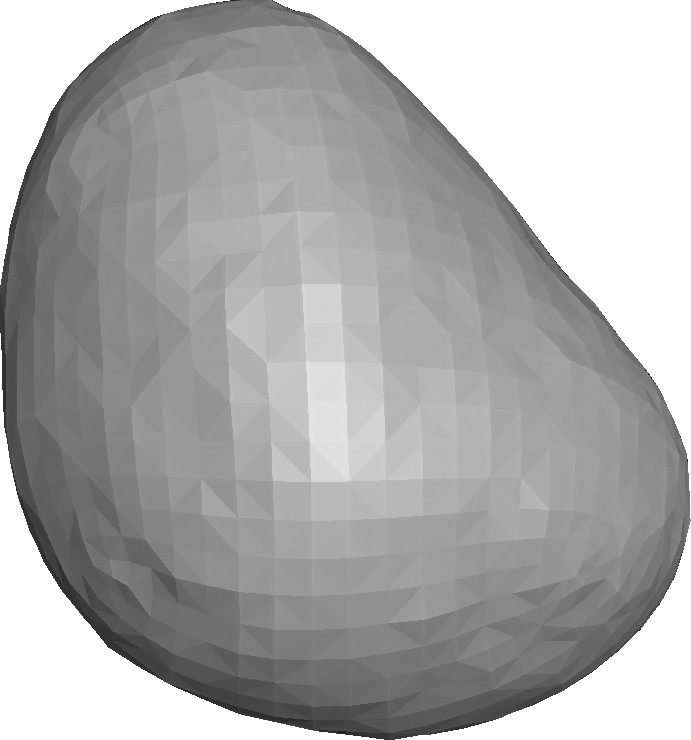}
    \includegraphics[width=0.30\linewidth]{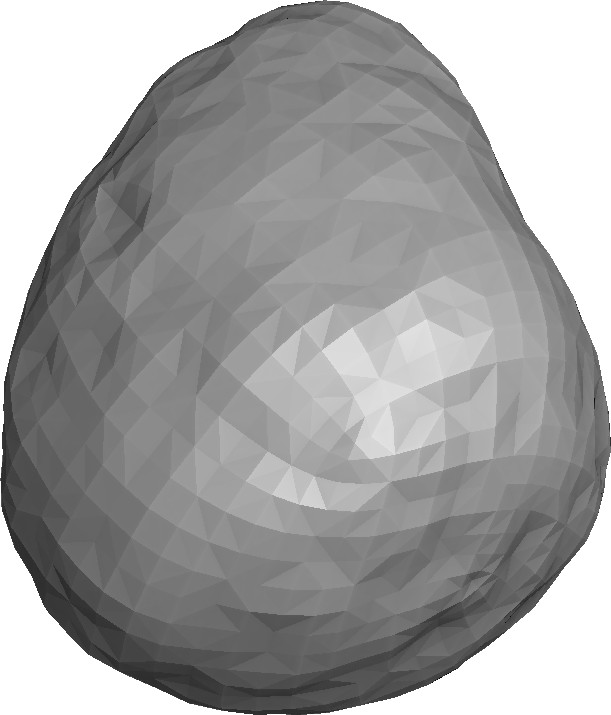}
    \includegraphics[width=0.30\linewidth,angle=90]{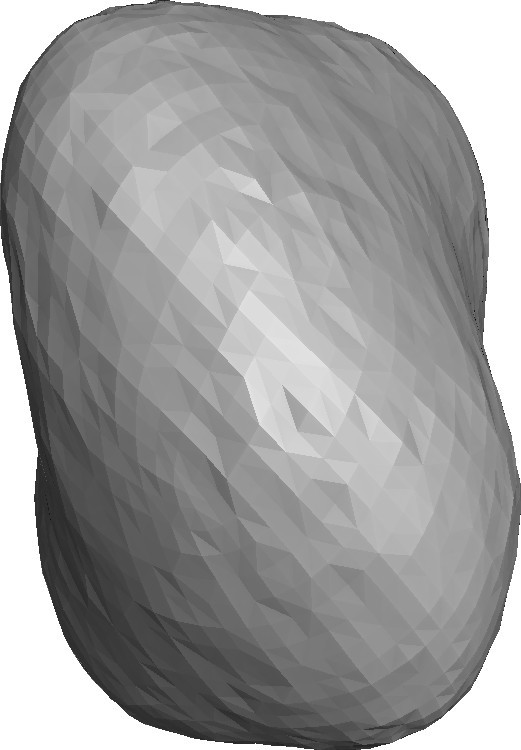}
    \includegraphics[width=0.30\linewidth,angle=90]{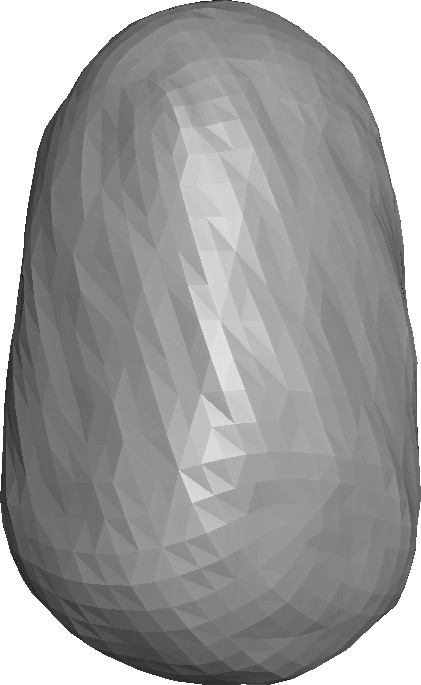}
    \vspace{2ex}
        \end{minipage}
      }
  \subfloat[structural problem definition]{
        \label{fig:teaser:problem}
        \begin{minipage}{0.2\linewidth}
          \centering
          \includegraphics[height=3.5cm]{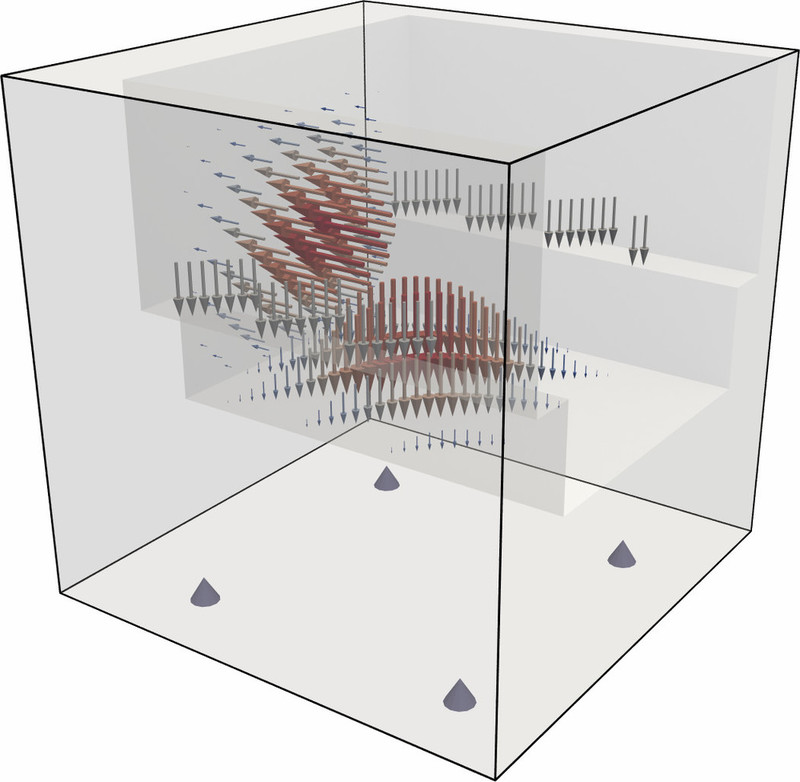}
        \end{minipage}
      }
  \subfloat[synthesized output]{
        \label{fig:teaser:simulation}
        \begin{minipage}{0.2\linewidth}
          \centering
          \includegraphics[height=3.5cm]{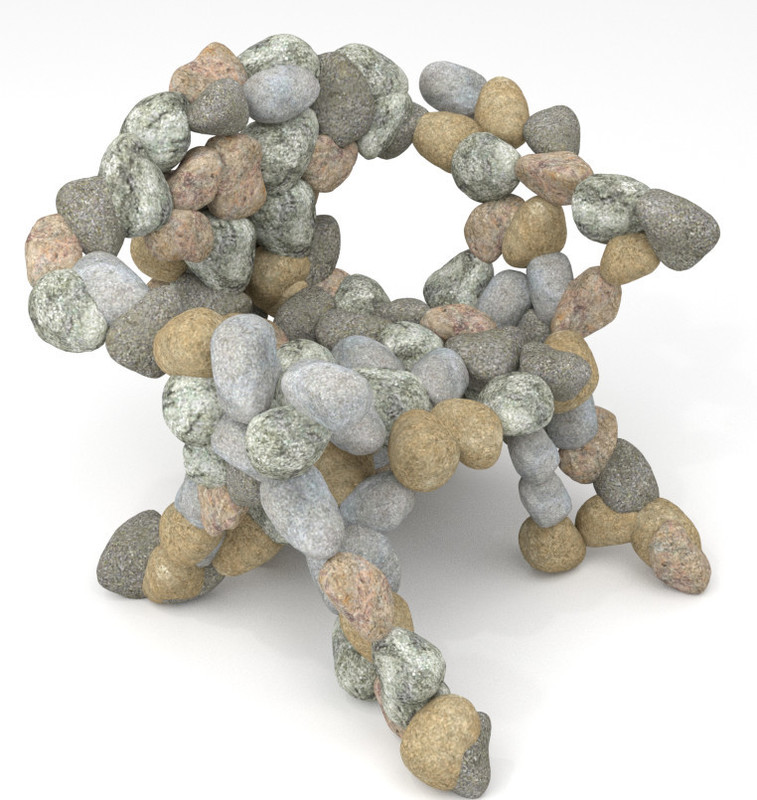} 
        \end{minipage}
      }
  \subfloat[manufactured output photographed in different views]{
        \label{fig:teaser:manufacture}
        \begin{minipage}{0.4\linewidth}
          \centering
          \includegraphics[height=3.5cm]{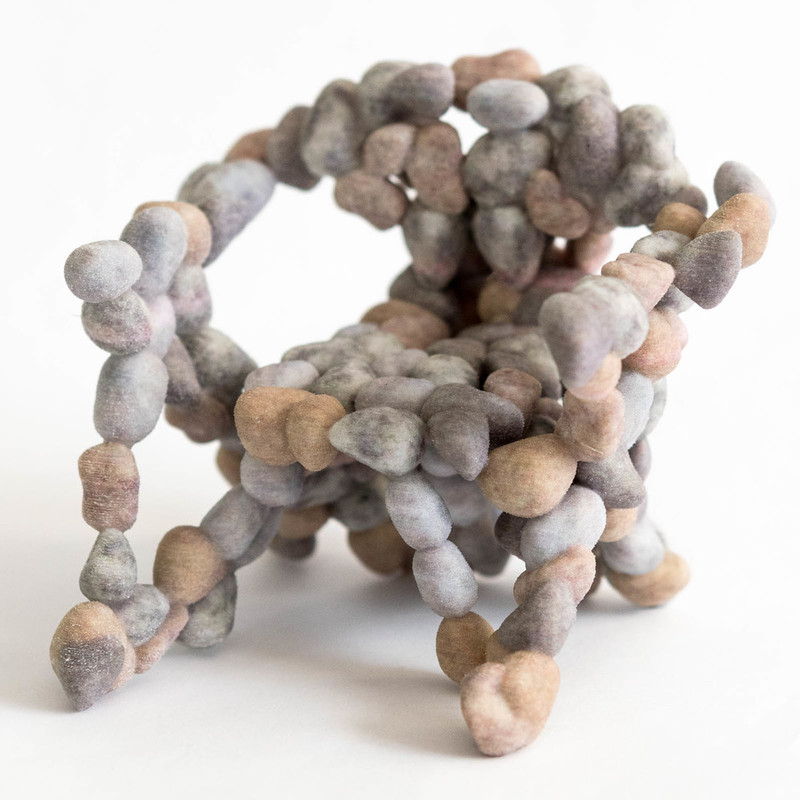}
          \includegraphics[height=3.5cm]{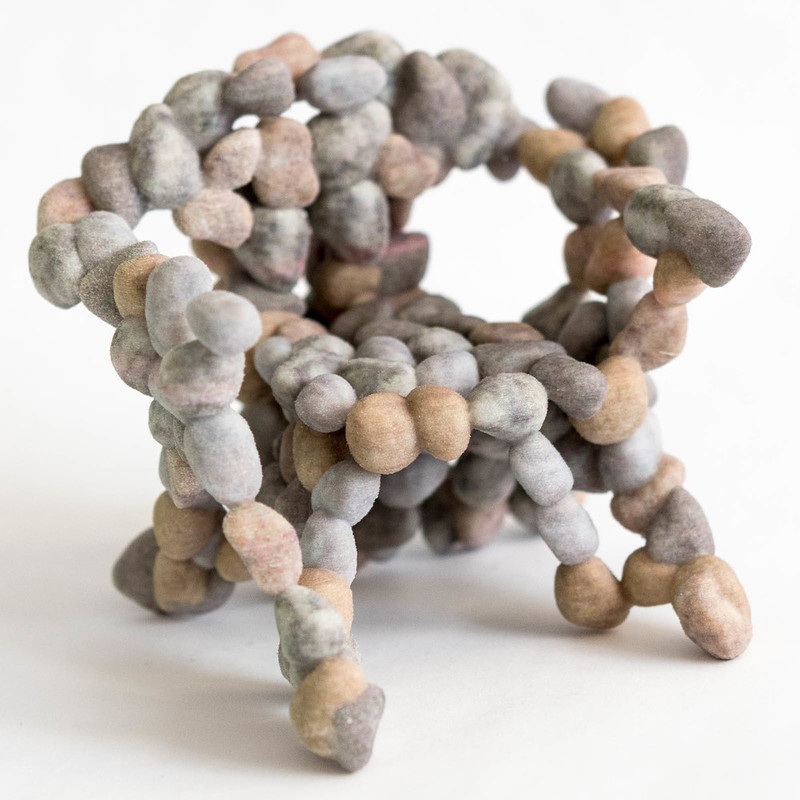}
        \end{minipage}
      }
  \Caption{Synthesizing element aggregates.}
      {%
        From the input elements \subref{fig:teaser:exemplar} and structural problem definition \subref{fig:teaser:problem}, our method automatically optimizes the output \subref{fig:teaser:simulation} that can be 3D printed \subref{fig:teaser:manufacture}.
	\subref{fig:teaser:problem} includes the loading scenario $-$ 4 anchors on the floor for chair legs and external forces visualized as vectors acting on the chair back and seat, which are visualized as transparent surfaces. The chair, including the legs, appeared spontaneously from the optimization process without prescribed shapes.
      }
  \label{fig:teaser}
  \label{fig:result:pebble_chair}
\end{teaserfigure}

%% file: body/10-introduction.tex
\section{Introduction}
\label{sec:introdution}

Depicting shapes by aggregating elements is a way for artists and designer to convey a specific look and feel, for instance creating furniture from wooden sticks~\footnote{\url{http://www.marvelbuilding.com/unique-wooden-chair-stacked-sticks-crossed-stick-chair.html}}, ropes \footnote{\url{https://www.dezeen.com/2014/12/22/a-zdvent-calendar-vermelha-chair-campana-brothers/}}, or metal ornaments \footnote{\url{https://www.dezeen.com/2013/05/08/brazilian-baroque-exhibition-by-the-campana-brothers/}}.
This is often used to create a striking contrast between the appearance of the base elements and the purpose of the object formed by their aggregation. Famous examples are the candy house depicted in Hansel and Gretel, the iron throne (made of swords) of the Game of Thrones series,  the paintings by Giuseppe Arcimboldo (portraits made of e.g. fruits, books, flowers), and the \enquote{accumulations} sculptures by artist Arman.

Aggregate elements are also ubiquitous in natural settings, e.g. a pile of rocks, a stash of fruits, a stone bridge.


Our goal is to allow the design of aggregate objects that can be manufactured and used in the real world. Challenges occur at two different scales: locally the elements have to be in contact with each others, but without significant overlaps that would destroy their appearance. Globally the elements have to form a rigid network so that the object is structurally sound.

Despite significant research on aggregates (see~\Cref{sec:related-work}), no solution currently exists for our purpose.
Available methods either optimize the manual assembly of specific 3D elements \cite{Yoshida:2015:ASH,Luo:2015:LOL}, or apply only to 2D domains such as planes or surfaces \cite{Dumas:2015:BES,Martinez:2015:SAA,Chen:2016:SOF,Zehnder:2016:DSS,Schumacher:2016:SDS}, or pack elements to closely approximate a target shape~\cite{Gal:2007:CEN}.

We present the first method to automatically generate 3D aggregates of elements arranged into globally rigid objects. We draw inspiration from modeling techniques that use rigidity as a design tool, by relying on topology optimization \cite{Christiansen:2015:CST,Wu:2016:ASF}. This lets the user specify the target shape through its desired structural properties.
Our system thus requires as input only a set of 3D elements and a set of constraints: the output domain, a loading scenario (forces and attachments), and optionally spatial constraints (volumes to be avoided). The output of our system is a physically sound object that can be 3D printed.
Our method is versatile and supports elements with shapes which are convex, concave, rigid or deformable.

Our algorithm optimizes aggregates by formulating a topology optimization problem with the position and orientation of individual elements as design variables.
Each element contributes density within a material grid used for structural analysis. We maximize global rigidity through a dedicated gradient descent scheme that relates elements degrees of freedom to local material densities.

Our setting is made challenging by the 3D nature of the problem and the number of elements.
The resolution of the simulation grid cannot be too high for reasonable speed and memory consumption.
To achieve proper convergence while maintaining performance we rely on a {\em continuation method} that gradually reduces the individual influence of element samples over the density grid, and a {\em connection step} to eliminate small gaps that might be filtered away by the grid resolution.

\input{body/11-contribution}

%% file: body/11-contribution.tex
\noindent The main contributions of this paper include:
\begin{itemize}
\item
The first system that automatically synthesizes structurally sound 3D aggregates from user-specified elements, structural objectives, and output domain constraints;

\item
The formulation of a topology optimization problem to optimize the positions and orientations of elements;

\item
The parameterization of element degrees of freedom for the support of convex, concave, rigid, as well as flexible elements of arbitrary shapes;

\item
A gradient-descent based solver that links output material densities to input element degrees of freedom through a differentiable density field function of elements' shapes;

\item A continuation method and a connection step to ensure convergence and quality while using coarse simulation grids for performance. 

\end{itemize}

We verify our method through both numerical simulation and physical manufacturing of a variety of objects with different shapes and elements.

%% file: body/20-related-work.tex
\section{Related Work}
\label{sec:related-work}

Aggregate geometry has been extensively explored in a variety of fields in graphics, engineering, and manufacturing. We focus below on the works most relevant to ours.

\paragraph{Graphics}

Aggregate geometry is ubiquitous and yet complex, and thus has been a research focus for modeling, animation, and rendering in computer graphics.
Such repetitive geometry is often too complex for manual authoring
and is better suited for procedural generation~\cite{Peytavie:2009:CGF,Guerin:2016:CGF}, data-driven synthesis from exemplar elements \cite{Sakurai:2014:MON} and their distributions \cite{Gal:2007:CEN,Ma:2011:DET,Landes:2013:ASA,Roveri:2015:EBR}. While some of these techniques fill an existing shape by packing elements inside, they cannot fully synthesize a rigid shape having only desired structural properties as input.

Physical simulation often have to deal with the dynamics of aggregated shapes, which are computationally demanding and hard to control. Nailing specific aspects such as contact, friction, and gravity among a subset of elements can greatly help stability, control, and speed \cite{Kaufman:2008:SPF,Hsu:2010:POO,Ritchie:2015:GDS}.
To reduce aliasing and increase rendering speed, \cite{Cook:2007:SSO} introduced a level-of-detail method to render aggregate geometry.


\paragraph{Design for manufacturing}
Large objects can be manually assembled from specific base elements, e.g. sticks \cite{Yoshida:2015:ASH} or LEGO bricks \cite{Luo:2015:LOL}.
These methods focus on computational support for manufacturing large, stable assemblies; they do not optimize the target shape itself and do not allow the user to choose the base elements.

Other approaches take into account both appearance and structure. \textcite{Zehnder:2016:DSS} focus on interactive editing, allowing the user to assemble curve elements into a fabricable connected network along a surface. The system automatically optimizes for deformations while indicating potential weak regions in the pattern.
Several methods formulate a structure optimization problem combing user-specified loading scenarios and exemplar patterns  \cite{Dumas:2015:BES,Martinez:2015:SAA,Chen:2016:SOF}.
These methods have been applied to 2D domains only, either planes or surfaces, but not to 3D volumes due to fundamental limitations in algorithms (e.g. \cite{Dumas:2015:BES,Chen:2016:SOF} are inherently 2D) or practical numerical feasibility (e.g. \cite{Martinez:2015:SAA} is computationally challenging for volumes).

The approach of \textcite{Schumacher:2016:SDS} covers a surface with holes having user specified shapes.
The layout is optimized for both structural properties and aesthetics, and is coupled with the structural simulation through the rigidity of the surface triangles.
Our method focuses on the complementary problem of optimizing a layout of elements that are in contact and form a globally rigid structure, in 3D, and using concave as well as deformable elements. 

Note that most of the aforementioned techniques attempt to \textit{cover} a domain with a pattern, while our technique, like \cite{Martinez:2015:SAA}, attempts to \textit{completely synthesize} a novel shape from structural and appearance constraints -- in our case the base elements forming the structure itself.

\paragraph{Mechanical engineering}
The problem of distributing solid elements or holes has been investigated in the field of mechanical engineering, as surveyed in~\cite{Lazarov:2016:LSA}.

There is a strong interest for jointly optimizing the position of embedded components and their support frame~\cite{Zhang:2011:SRA,Zhang:2012:ILD,Kang:2013:ITO,Xia:2013:AIM,Wang:2014:TDO,Zhang:2015:ELC}.
%
However, approaches that produce results comprised of aggregated elements only, i.e. without embedding them into an optimized support frame, have recently gained interest. These approaches relate the elements to an underlying simulation in a material grid, considering that each element generates a local density field~\cite{Overvelde:2012:TMN,Guo:2014:DTO,Norato:2015:AGP,Zhang:2016:ANT,Deng:2016:DFS}.
Avoiding overlaps between elements can be challenging. Explicit constraints can be used~\cite{Zhang:2015:ELC,Gao:2015:AIA,Kang:2016:STO}, or elements can be combined by differentiable CSG~\cite{Chen:2007:SOW,Chen:2008:SSO,Liu:2014:ALS} using r-functions~\cite{Rvachev:1982:TOR} or max operators relaxed as $p$-norms~\cite{Norato:2015:AGP}. When applicable, the density field around each element can be modified to discourage overlaps~\cite{Overvelde:2012:TMN,Guest:2015:OTL}.

Our approach is inspired by 2D methods using density fields around elements~\cite{Overvelde:2012:TMN,Norato:2015:AGP}. However, contrary to these techniques, we consider a 3D formulation and use elements that can have arbitrary shapes (elongated, concave, flexible), while these approaches typically focus on simpler, parametric shapes. We also target larger number of elements: the aforementioned techniques optimize for a few elements, while our smallest results have a hundred of them.

This makes convergence more challenging, and increases computational costs significantly.
Besides the 3D formulation and parameterization of the element's degrees of freedom, we introduce a continuation scheme on the element densities to allow for an improved convergence within reasonable computation times, as well as a connectivity improvement step.

%% file: body/30-method.tex
\section{Overview}
\label{sec:method}

Our algorithm takes as input a set of elements as colored 3D meshes, an output domain (size and shape), and a description of the desired object structural's role: external loads to be supported as well as fixed attachment points. This is illustrated in \Cref{fig:teaser:exemplar,fig:teaser:problem}.
Based on the exemplar elements and output domain, an initial configuration is created by distributing an estimated number of random instances from the exemplar elements.

Our optimizer iterates from this initial configuration to produce a rigid shape composed of the base elements, see \Cref{fig:teaser:simulation}. Since the aggregated elements form a single connected structure optimized for rigidity, the object can be 3D printed and used as intended, see \Cref{fig:teaser:manufacture}.

The elements in the domain are defined by a set of parameters~$\allDofs$, which are the design variables used in the optimization.
The structural objective we minimize is the compliance $\compliance$ of the system, which can be understood as the inverse of the rigidy. It is computed from the current shape configuration $\allDofs$ and the user specified external forces $\externalForces$.
We thus seek to solve the following constrained optimization problem with respect to the set of parameters $\allDofs$:
\begin{subequations}
	\label{eq:optimization-problem-def}
	\begin{align}
		\minimize_\allDofs \quad & \complianceFunc{\allDofs}
		\label{eq:optim:objective-function}\\
		\text{subject to} \quad & \domainFunction(\allDofs) \leq 0
		\label{eq:optim:domain-constraint}\\
		& \allDofs_{\min} \leq \allDofs \leq \allDofs_{\max}
		\label{eq:optim:box-constraints}
	\end{align}
\end{subequations}

where
the constraint $\domainFunction$ forces elements to stay confined to a user-defined output domain, while $\allDofs_{\min}$ and $\allDofs_{\max}$
are box constraints on the element parameters $\allDofs$, which are used to limit deformations on flexible elements and retain element's centroids within the domain.

Due to the large number of design variables we seek to use a gradient-based method. To do so, we have to define the
compliance $\compliance$ and its gradient in terms of the element parameters~$\allDofs$.

\input{figs/pipeline}

The difficulty is that computing the compliance requires an underlying finite element simulation. Using the elements directly would impose an expensive union and tetrahedral remeshing step at every iteration. Instead, we build upon density-based topology optimization methods~\cite{Bendsoe:2004:TOT}. These approaches represent the optimized shape as a 3D density grid $\allDensities$, where a density of $1$ is solid material and $0$ empty.
A \emph{soft} elastic material is assigned to the grid cells with density $\densityGridCell = 0$, and a \emph{rigid} solid material is assigned to the grid cells with density $\densityGridCell = 1$. In between, the material stiffness is interpolated according to $\densityGridCell$.
Rather than directly manipulating the grid cell densities as design variables, we define the cell densities from the elements and their parameters $\allDofs$. Intuitively, each element fills  with solid material the density grid cells it overlaps, as illustrated in~\Cref{fig:pipeline}. Given a differentiable density field function for the elements' shapes, it is then possible to compute the compliance gradient through the chain rule.

However, using arbitrary shapes that optionally can deform makes finding a differentiable density field function challenging. We propose to approximate the geometry of the elements with multiple samples, each supporting a density kernel based on a differentiable smoothed Heaviside function. When the elements are allowed to deform, the samples are allowed a limited range of relative motions.

We next detail our problem formulation (\Cref{sec:formulation}), and the numerical scheme we design to follow the compliance gradient and maximize the system's rigidity (\Cref{sec:solver}).

\section{Problem formulation}
\label{sec:formulation}


In the following, we first discuss our choice of parameterization for elements in 3D, for rigid and flexible elements (\Cref{subsec:method:element-parameterization}).
We next detail how we express domain constraints $\domainFunction$ (\Cref{subsec:constraints}) and how the densities and their gradients are computed from the point samples (\Cref{subsec:densities}). Finally we describe the formula for computing the compliance and its derivative (\Cref{subsec:compliance}).
For convenience, we give a table with all notations in the supplemental material.

\subsection{Element Parameterization}
\label{subsec:method:element-parameterization}

We represent the elements in the domain by point samples~\cite{Ma:2011:DET}.
The samples of an element can be selected manually, or they can be computed automatically. We chose the latter option, and used a centroidal Voronoi tessellation \cite{Liu:2009:OCV} to sample points regularly inside the input meshes (see \Cref{fig:pipeline}, left).
Each sample is associated with a radius $\sampleRadius$, that can vary from sample to sample, e.g. depending on how well the ball of center $\samplePosition$ and radius $\sampleRadius$ locally approximates the shape of the input element.

We denote the set of all samples as $\allSamples$, and their world-space position by the matrix $\allSamplePos \in \reals^{3 \times \size{\allSamples}}$.

\paragraph{Rigid elements}

When the user chooses to preserve the element's shape during the optimization, they undergo only a rigid transformation parameterized by a translation and a rotation.
For algorithm simplicity and computation efficiency, we do not allow elements to rescale, which can be achieved by providing input elements with different sizes.

Consider an element $\element$ in the output domain. The element is defined by a set of parameters $\allDofsElement = (\dofsTranslation, \dofsRotation, \dofsSamplePos)$, where $\dofsTranslation$ denotes the translational degrees of freedom, $\dofsRotation$ the rotational dofs, and $\dofsSamplePos$ the individual positions of the samples before transformation.
The positions of the element samples in the output domain are given by the following relation:
\begin{align}
	\allSamplePosElement(\allDofs) = \elementRotation(\allDofs) \elementLinear \allDofsSamplePosElement(\allDofs) + \elementTranslation(\allDofs)
\end{align}
, where $\elementRotation(\allDofs)$ is the rotation matrix which is a function of the rotation parameters $\dofsRotation$ (exponential map in 3D, details in \Cref{subsec:rotparam}), $\elementLinear$ is a fixed linear transformation (fixed rotation + scaling), $\elementTranslation(\allDofs)$ is the translation matrix of the element centroid, and depends only on $\dofsTranslation$. Finally, $\allDofsSamplePosElement(\allDofs)$ are the (untransformed) sample positions. For rigid elements, $\allDofsSamplePosElement$ is constant and determined once for each element type, while for elements allowed to deform $\allDofsSamplePosElement(\allDofs)$ is a matrix representing all the positional dofs $\dofsSamplePos$ for element $\element$.

\input{body/39-deformable}

\input{body/33-constraints}

\subsection{Material Densities}
\label{subsec:densities}

To compute the compliance of the shape being synthesized, it is necessary to assign a density everywhere in space from the current configuration of the point samples.

\paragraph{Representation}

Let $\sample$ be a point sample in the output domain, and let $\densitySampleFunc$ be the material density associated to the sample $\sample$.
We define $\densitySampleFunc \colon \reals^3 \to \reals_+$ as a radial basis function, which depends only on the distance from the sample position $\samplePosition$, and is parameterized by the sample radius~$\sampleRadius$.
The RBF $\densitySampleFunc$ is chosen to have a compact support so that each element only has a local influence.
In practice, we use a smoothed Heaviside step function:
\begin{align}
	\densitySampleFunc(\currentPosition) =
        \densityElementFunc(\currentPosition)
        \left(
	\frac{1}{2} + \frac{1}{2} \tanh\left( \rbfSlope \left(
		\sampleRadius^2
		- \left(\frac{\norm{\currentPosition - \samplePosition}}{\rbfWidth}\right)^2
	\right) \right)
        \right)
	\label{eq:material-densities:heaviside}
\end{align}
where $\rbfSlope$ controls the smoothness of the approximation, and $\rbfWidth$ controls the radius of influence.
In our implementation we consider that $\densitySampleFunc(\currentPosition) = 0$ when $\norm{\currentPosition - \samplePosition} \geq 3 \rbfWidth \sampleRadius$.
The term $\densityElementFunc$ considers whether $\currentPosition$ is inside the element $\element \ni \samplePosition$ so that small gaps will not be erroneously filled during FEM in \Cref{subsec:compliance}:
\begin{align}
\densityElementFunc(\currentPosition) = \insideElement{\currentPosition}
\label{eq:density_element}
\end{align}

In contrast to works in mechanical engineering that use elements for structural optimization (see \Cref{sec:related-work}), in \Cref{eq:material-densities:heaviside} multiple samples correspond to a same element, and are driven by the same element's parameters (a rigid transformation, or a skeletal parameterization if deformable). In addition, we propose to use  $\rbfSlope$ and $\rbfWidth$ in a continuation scheme during optimization to improve convergence, as will be detailed in \Cref{subsec:continuation}.

\paragraph{Gathering}

The total material density $\densityAllSamples(\currentPosition)$ at any point in $\reals^3$ is defined as the max of the densities induced by all the sample points $\sample \in \allSamples$:
\begin{align}
	\densityAllSamples(\currentPosition) = \max_{\sample \in \allSamples}(\densitySampleFunc(\currentPosition))
	\label{eq:material-densities:max}
\end{align}

Defining the material density using a maximum instead of a sum discourages elements to overlap. Indeed, overlapping two elements under a max decreases the total density of the system, which increases the compliance. Thus, solutions where samples do not overlap have a lower compliance and are preferred. Conversely, minimizing the compliance tends to pull the elements closer together, as inter--element gaps result in fragile structures. This is precisely the behaviour we intend: pulling the elements together while discouraging large overlaps.

It should be noted that the $\max$ function is technically not differentiable.
A common workaround is to resort to a smoothed $\max$ formulation, e.g. using a $p$-norm $\norm{\cdot}_p$, with a high $p \geq 6$ or $8$.
The drawback of this approach is that the actual density at any point in space depends on the number of samples $\size{\allSamples}$, as the $p$-norm inevitably computes a form of weighted average over the domain. To retrieve a good approximation of the $\max$ function, it is necessary to increase the $p$ exponent when there are a high number of samples, which leads to numerical inaccuracies.

In practice, we have found that simply ignoring this theoretical issue and retaining a hard $\max$ formulation does not impede the overall gradient computation.
Indeed, the only non-differentiable points of the $\max$ function are the points which are closest and equidistant from two different samples (assuming $\sampleRadius$ is the same for all samples).
In other words, they are located on the edges of the Voronoi diagram formed by the current distribution of samples. Due to the limited numerical precision of floating point operations these singular points never occur in practice.

\paragraph{Discretization}

In order to compute the compliance of the system via a finite element method, the material densities are discretized in a regular 3D grid, in our case composed of linear H8 cube elements.
Let $\gridCell$ denote a cell in the regular grid $\allGridCells$, and let $\densityGridCell$ be its associated material density. The discretized cell density is given by the following relation:

\begin{align}
	\densityGridCell = {
		\frac{1}{\volumeGridCell}
		\int_{\currentPosition \in \gridCellDomain} \densityAllSamples(\currentPosition)
		\dd{\currentPosition}
	}
	\label{eq:material-densities:grid-cell-int}
\end{align}
where $\gridCellDomain$ is the domain covered by the grid cell $\gridCell$.

\Cref{eq:material-densities:grid-cell-int} simply means that $\densityGridCell$ is defined as the average density $\densityAllSamples$ over the grid cell $\gridCell$.
In practice, the integral \eqref{eq:material-densities:grid-cell-int} can be computed either 1) analytically by an exact expression of the integral of $\densityAllSamples(\currentPosition)$, or 2) numerically by means of a Gaussian quadrature rule, i.e. evaluating $\densityAllSamples(\currentPosition)$ at specified points inside the cell.
While the expression of $\densitySampleFunc(\currentPosition)$ for a given sample $\sample$ is integrable analytically, the use of the $\max$ function makes it difficult to derive a simple analytic expression of the resulting integral.
For this reason, we opted for numerical integration of the expression given in \Cref{eq:material-densities:grid-cell-int}:

\begin{align}
	\densityGridCell = {
		\frac{1}{\volumeGridCell}
		\sum_{k=1}^N \gaussWeight_k \densityAllSamples(\gaussPoint_k^\gridCell)
	}
	\label{eq:material-densities:grid-cell-sum}
\end{align}
where $\gaussPoint_k^\gridCell$ is the $k$-th evaluation point of the quadrature rule for the cell $\gridCell$, and $\gaussWeight_k$ is the associated weight.
In our experiments, we used a 2-point quadrature rule (i.e. 8 points per cube).

\subsection{Compliance and Sensitivities}
\label{subsec:compliance}

Given the 3D density grid, and knowing the external forces $\externalForces$ applied to the system, we can now compute how the system deforms under load.
The discrete displacement field $\displacements$ is obtained with the finite element method, solving the equilibrium equation:
\begin{align}
	\label{eq:fem-equilibrium}
	\stiffnessMatrix(\allDensities) \displacements = \externalForces
\end{align}
where $\stiffnessMatrix(\allDensities) = \sum_{\gridCell \in \allGridCells} \densityGridCell \left[\begin{smallmatrix}\stiffnessMatrixBase\end{smallmatrix}\right]$ is the global stiffness matrix of the system, assembled from the individual matrices of every grid cell $\gridCell$, and where $\stiffnessMatrixBase$ is the stiffness matrix for the base solid material. We use a solver similar to Wu et al.~\shortcite{Wu:2016:ASF} to solve for \Cref{eq:fem-equilibrium}.

The compliance of the system is then computed as~\cite{Bendsoe:2004:TOT}:
\begin{align}
	\complianceFunc{\allDensities}
	= \transpose{\displacements} \externalForces
	= \sum_{\gridCell \in \allGridCells} \densityGridCell \transpose{\localDisplacements} \stiffnessMatrixBase \localDisplacements
\end{align}
where $\localDisplacements$ is the displacement vector associated to the node of the grid cell $\gridCell$ induced by the external force $\externalForces$.

Using the adjoint method (\cite{Bendsoe:2004:TOT}, \S 1.2.3), the partial derivatives of the compliance can be expressed as

\begin{align}
	\pderivative{\compliance}{\densityGridCell} =
	- \transpose{\localDisplacements} \stiffnessMatrixBase \localDisplacements
	\label{eq:compliance-pderiv}
\end{align}

\paragraph{Chain Rule}

The current pipeline for computing the compliance $\compliance$ from the element parameters can be summarized as follows:

\begin{align}
	\allDofs
	\stackrel{\dofsToSamplePos}{\to} \allSamplePos
	\stackrel{\samplePosToDensities}{\to} \allDensities
	\stackrel{\densitiesToScalar}{\to} \compliance
\end{align}
where $\allDofs$ denotes the element parameters, $\allSamplePos$ denotes the sample positions in the output domain, $\allDensities$ are the grid cell densities, and $\compliance$ is the scalar value of the compliance objective function.
\Cref{eq:compliance-pderiv} explains how to obtain the compliance gradient $\grad{\densitiesToScalar} \in \reals^{\size{\allDensities}}$.
Given $\grad{\densitiesToScalar}$, the gradients $\grad{(\densitiesToScalar \circ \samplePosToDensities)}$ and $\grad{(\densitiesToScalar \circ \samplePosToDensities \circ \dofsToSamplePos)}$ can be computed efficiently via the chain rule. If we note the gradient as a row-vector, this can be expressed as:
\begin{align}
	\grad{(\densitiesToScalar \circ \samplePosToDensities \circ \dofsToSamplePos)} =
		\grad{\densitiesToScalar} \;
		\jacobianMatrix{\samplePosToDensities} \;
		\jacobianMatrix{\dofsToSamplePos}
	\label{eq:chain-rule-jacobian}
\end{align}
Note that the Jacobian matrices $\jacobianMatrix{\samplePosToDensities}$ and $\jacobianMatrix{\dofsToSamplePos}$ are sparse matrices, so the products in \Cref{eq:chain-rule-jacobian} can be implemented efficiently.

In the following, we explain how to compute the Jacobian matrices $\jacobianMatrix{\samplePosToDensities}$ and $\jacobianMatrix{\dofsToSamplePos}$. The sensitivity of the cell density $\densityGridCell$ with respect to the $\sample$th-sample $j$th-coordinate is

\begin{align}
	\jacobianMatrix{\samplePosToDensities}[\gridCell, \sample_j]
	= {
		\pderivative{\densityGridCell}{\samplePositionCoord{j}}
	}
	= {
		\frac{1}{\volumeGridCell}
		\sum_{k=1}^N \gaussWeight_k
			\pderivative{
				\densityAllSamples(\gaussPoint_k^\gridCell)
			}{\samplePositionCoord{j}}
	}
\end{align}
with
\begin{align}
	\pderivative{
		\densityAllSamples(\gaussPoint_k^\gridCell)
	}{\samplePositionCoord{j}}
	= {
		\pderivative{
			\max_{\otherSample \in \allSamples}(\densityOtherSampleFunc(\gaussPoint_k^\gridCell))
		}{\samplePositionCoord{j}}
	}
	\label{eq:material-density-pderiv}
\end{align}

When the $\max(\cdot)$ in \Cref{eq:material-density-pderiv} is reached by a single sample $\sample^\star$, then partial derivative $\pderivative{\densityAllSamples(\gaussPoint_k^\gridCell)}{\samplePositionCoord{j}}$ exists, and is non-zero when $\sample^\star = \sample$.
It is then equal to the partial derivative $\pderivative{\densitySampleFunc(\gaussPoint_k^\gridCell)}{\samplePositionCoord{j}}$, whose expression is obtained by deriving \Cref{eq:material-densities:heaviside} (detailed in the supplemental material).

For the element positions, the partial derivative with respect to element parameters $\oneDof{k}$ can be expressed as:
\begin{align}
	\jacobianMatrix{\dofsToSamplePos}[\sample_j, k]
	= {
		\pderivative{\samplePositionCoord{j}}{\oneDof{k}}
	}
\end{align}

If sample $\sample$ belongs to element $\element$, then we can write:

\begin{align}
	\pderivative{\samplePosition}{\oneDof{k}} =
	\pderivative{\oneDof{k}}(
		\elementRotation \elementLinear \dofsSamplePos_\sample + \elementTranslation
	)
	\label{eq:sample-position-pderiv}
\end{align}

The expression of \Cref{eq:sample-position-pderiv} can be simplified whether $\oneDof{k}$ corresponds to the rotation parameter, the translation, or the sample positions. The different cases are presented in the supplemental.

Regarding deformable elements, in order to compute the gradient of the word-space sample position with respect to the sample parameters (\Cref{eq:sample-position-pderiv}), we need to be able to compute $\pderivative{\dofsSamplePos_\sample}{\oneDof{k}}$ when $\oneDof{k}$ is one of the $\{ \dofsAngularPos_i \}_i$ that parameterize the sample position $\dofsSamplePos_\sample$. The complete derivation of this partial derivative is also detailed in the supplemental material.

We now have all the definitions that form the basis of our formulation. We proceed with describing our numerical solver in~\Cref{sec:solver}.

%% file: figs/pipeline.tex
\begin{figure}[htb!]
	\centering

	\includegraphics[width=0.9\linewidth]{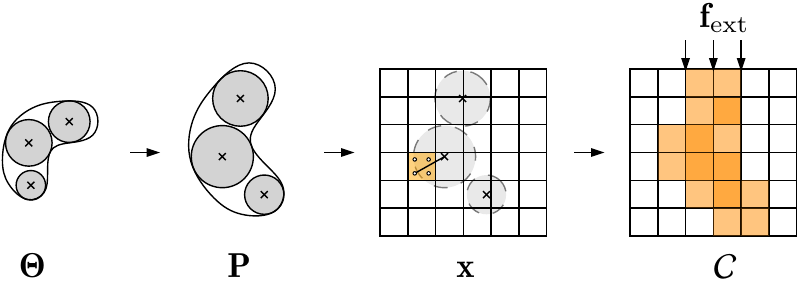}

	\Caption{Our method pipeline.}
        {%
          The element/sample parameters $\allDofs$ decide the element distribution $\allSamplePos$, from which we compute the material densities $\allDensities$ for compliance analysis $\compliance$.
        }
	\label{fig:pipeline}
\end{figure}

%% file: body/39-deformable.tex
\paragraph{Deformable elements}

Optionally, the user may choose to allow elements to deform during the optimization process.
In addition to the rigid degree of freedom $(\dofsRotation, \dofsTranslation)$ associated to an element $\element$, we use the sampled points inside each element (coordinates in the local element reference frame) as control points to drive a non-rigid deformation.
At the end of the process we use ARAP \cite{Sorkine:2007:ASM} and the implementation from \libigl{} \cite{Jacobson:2016:IGL} to recover the final shape.

To prevent excessive deformation during the optimization, we limit how much the control points can stray from the initial configuration.
One approach would be to add a set of constraints on the positions of the constituent samples $\dofsSamplePos$ in \Cref{eq:optimization-problem-def}.
However, we have found that adding such non-linear constraints makes the optimization problem too challenging to solve: the solver either tends to get stuck (constraints too tight), or produces excessive deformations (constraints too loose), without striking a good balance.


\input{figs/deformable-parameterization}

Instead we propose to rely on a parameterization of the point sample positions that eliminates the need for complex constraints.
The parameterization we propose is described in \Cref{fig:deformable-parameterization}.
We observe that most often, the geometry of deformable elements can be well captured by a skeleton. We thus embed an articulated skeleton within the element's shape. The lengths of the skeleton's bones remain constant, preventing longitudinal distortions. The angles can be restricted through simple box constraints (see $\allDofs_{\min}, \allDofs_{\max}$ in \Cref{eq:optimization-problem-def}) to prevent excessive rotations at joints.

Let us assume that we are given a skeleton (tree) $\rootedTreeElement$ connecting the samples within an element $\element$.
(We describe an algorithm for the automatic construction of $\rootedTreeElement$ in the supplemental material.)
We define the position of each sample relative to its parent, parameterized by a rotation $\rotationMatrix(\dofsAngularPos_\sample)$:
\begin{align}
	\dofsSamplePos_\sample \eqdef \rotationMatrix(\dofsAngularPos_\sample) \deltaSamplePos_\sample + \dofsSamplePos_{\parent{\sample}}
        \label{eqn:dof_sample_pos_deform}
\end{align}
Note that $\deltaSamplePos_\sample$ is defined based on the input, undeformed element, but is fixed through the optimization. The root of the tree, $\dofsSamplePos_0$, is also fixed/constant through the optimization (without loss of generality since each element has already a translational degree of freedom).
In the undeformed configuration, $\dofsAngularPos_\sample = 0$ for all samples and the reconstructed surface is the same as input element.
\Cref{fig:results:noodles} shows a result where different limits are set for the acceptable deformations.

One limitation of the skeleton approach, however, is that we are restricted to a set of skeleton--like elements and cannot directly support deformable elements such as sheets or flexible rings.

%% file: figs/deformable-parameterization.tex
\begin{figure}[b]
	\centering

	\includegraphics{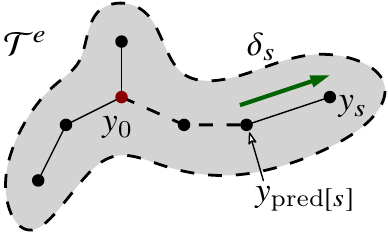}

	\Caption{Deformable element parameterization.}
        {%
        The position of a sample dof is expressed hierarchically as in \Cref{eqn:dof_sample_pos_deform}. The root sample position is fixed in the local coordinate system of the element.
	}
	\label{fig:deformable-parameterization}
\end{figure}

%% file: body/33-constraints.tex
\subsection{Domain Constraints}
\label{subsec:constraints}



Similarly to \cite{Ma:2011:DET}, we seek to restrict the sample positions to the interior of a user-given boundary. In addition, each sample $\sample$ must be at a distance at least $\sampleRadius$ from the boundary surface, otherwise part of the element will fall outside the domain.
Let $\distanceToBoundary{\currentPosition}$ be the signed distance from a point to the domain boundary, with negative values inside.
We define the following functional:
\begin{align}
	\domainFunction(\allSamplePos) =
		\sum_{\sample \in \allSamples} \max(0, \distanceToBoundary{\samplePosition} - \sampleRadius)
\label{eqn:domain_boundary}
\end{align}
It follows that we have $\domainFunction(\allSamplePos) \leq 0$ if and only if all the samples lie inside the domain at a distance at least $\sampleRadius$.

We always activate the boundary constraint for the domain. Optionally, it can be used to forbid regions of the output domain, e.g. the surfaces along chair back and seat in \Cref{fig:result:pebble_chair}, for seating comfort.

%% file: body/35-implementation.tex
\section{Solver}
\label{sec:solver}

The problem defined in \Cref{eq:optimization-problem-def} is non-linear and non-convex, with constraints that are also non-linear and non-convex. In addition, computing the objective function involves an expensive FEM computation to compute the equilibrium state (\Cref{eq:fem-equilibrium}).
A solver of choice for solving topology optimization problems is MMA \cite{Svanberg:1987:TMO}, a gradient-based optimizer, which can handle such non-linear constraints and does not require a line-search at every update step (thus avoiding solving the expensive \Cref{eq:fem-equilibrium}).
In practice, we use a custom implementation of MMA in C which allows us to use the continuation method mentioned in \Cref{subsec:continuation}

The complete pseudo code is given in \Cref{alg:optimizer-main-loop}.

\input{supplemental/00-code}

\input{body/31-init}

\input{body/45-continuation-fig}

\subsection{Iterations}
\label{subsec:method:iter}

After initialization the optimizer enters an iterative loop that computes the compliance gradient with respect to element parameters (\Cref{subsec:compliance}) and performs a descent step, following the MMA method.
We however embed two additional mechanisms. The first is a continuation scheme that affords for improved convergence in our setting (\Cref{subsec:continuation}).
The other is a connectivity step that explicitly encourages connections between neighboring elements  (\Cref{subsec:connectivity}).
Finally, we provide some details on how 3D rotations are handled during optimization (\Cref{subsec:rotparam}).

The total number of iterations is determined by the continuation scheme (see \Cref{subsec:continuation}, and pseudo-code \Cref{alg:optimizer-main-loop}).

\input{body/34-continuation}

\input{body/38-connect}

\input{body/37-rotparam}

%% file: supplemental/00-code.tex
\vspace*{-1ex}
\begin{algorithm}[htbp]
\begin{minted}[linenos, frame=none, framesep=0mm, numbersep=1mm, mathescape]{lua}
function optimize()
	local num_iter = 30;
	local alpha = 3.0;
	local beta = 1.0;
	local alpha_min = 0.9;
	local beta_max = 2.0;
	for k = 1, num_iter do
		-- Main MMA update
		Update();
	end
	while alpha > alpha_min do
		alpha = math.max(alpha_min, alpha * 0.9);
		SetContinuationParameters(alpha, beta);
		ReparameterizeRotations();
		for k = 1, num_iter do
			-- Main MMA update
			Update();
		end
		if alpha < 2.0 then
			-- Sub-solver step: 10 BGFS updates
			BuildConnectivityGraph(); -- Described in $\text{\Cref{alg:build-connectivity-graph}} \label{line:build-connectivity-graph}$
			OptimizeConnectivity(10);
		end
	end
	while beta < beta_max do
		beta = math.max(beta_max, beta * 2.0);
		SetContinuationParameters(alpha, beta);
		ReparameterizeRotations();
		for k = 1, num_iter do
			-- Main MMA update
			Update();
		end
		if alpha < 2.0 then
			-- Sub-solver step: 10 BGFS updates
			BuildConnectivityGraph();
			OptimizeConnectivity(10);
		end
	end
end
\end{minted}
\caption{Pseudo-code of our solver.}
\label{alg:optimizer-main-loop}
\end{algorithm}
\vspace*{-1ex}

%% file: body/31-init.tex
\subsection{Initialization}
\label{subsec:method:initialization}

\paragraph{Number of elements/samples}

For efficiency reasons we maintain the number of elements/samples constant during the main optimization loop: Adding or removing elements would reset the gradient information accumulated by the MMA optimizer over the previous iterations.

The user can thus specify the desired number of elements directly, or indirectly through a ratio of the output volume that should be occupied by elements (the overall volume does not change during optimization as the number of elements is kept constant and deformation constraints limit volume increase).

\paragraph{Placing elements/samples}

We randomly place the initial elements and spread them out by computing a centroidal Voronoi tessellation (CVT) of their centers. As this ignores the actual shape of the elements, some may lie partially outside the domain.
We improve the initial distribution by next minimizing the CVT objective on the \emph{samples} composing the elements, under the boundary constraint (\Cref{eqn:domain_boundary}). This pulls all samples back inside the output domain boundary, while preserving a uniform distribution.
An illustration of the resulting initial distribution is presented in \Cref{fig:numiteration}.

%% file: body/45-continuation-fig.tex
\begin{figure*}[t]
  \centering
  \subfloat[no continuation nor connectivity]{
    \label{fig:continuation:none}
    \includegraphics[width=0.32\linewidth]{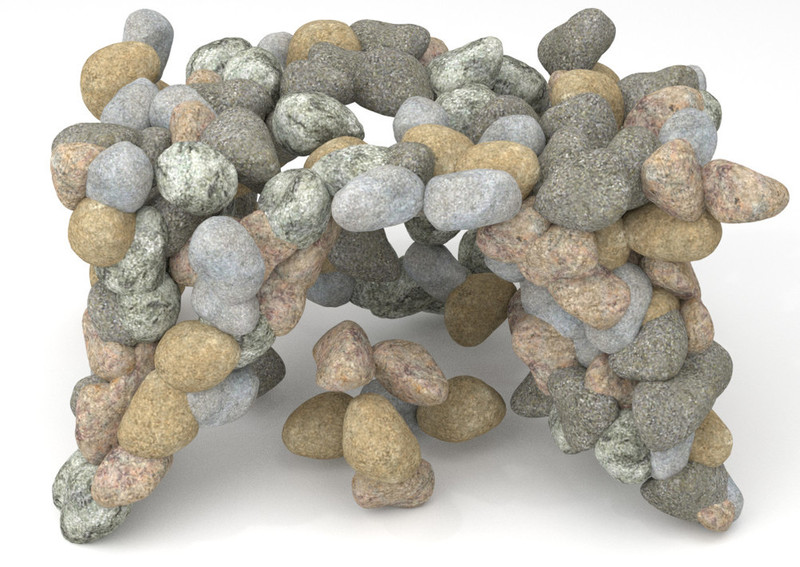}
  }
  \subfloat[with continuation but no connectivity]{
    \label{fig:continuation:no_connectivity}
    \includegraphics[width=0.32\linewidth]{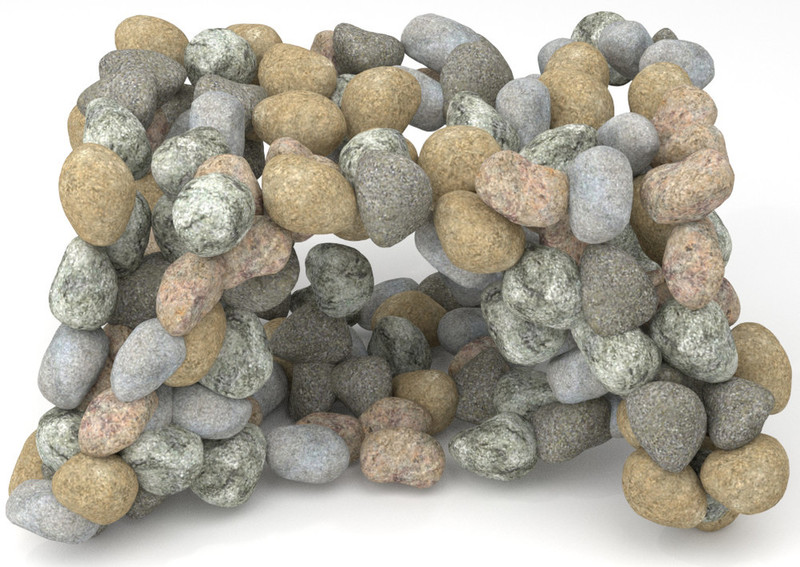}
  }
  \subfloat[with continuation and connectivity]{
    \label{fig:continuation:both}
    \includegraphics[width=0.32\linewidth]{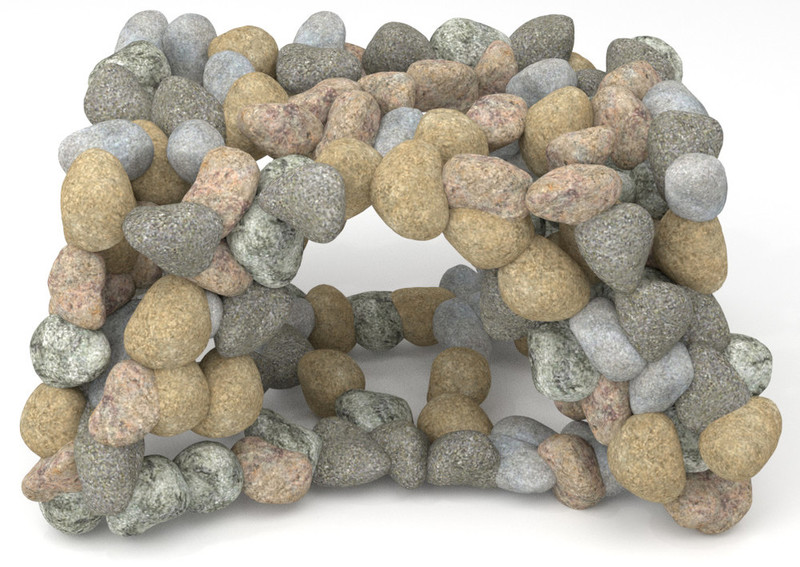}
  }
\vspace*{-3mm}
  \Caption{The effects of continuation and connectivity improvement steps.}
  {%
Our continuation method helps avoid stranded elements, while connectivity improvement reduces gaps between elements.
\subref{fig:continuation:none} With neither continuation nor connectivity some elements tend to isolate from the rest.
\subref{fig:continuation:no_connectivity} With continuation only the results improve but some elements tend to stretch out in different directions.
\subref{fig:continuation:both} Enabling both continuation and connectivity resolves those problems.
  }
  \label{fig:continuation}
    \vspace*{-2mm}
\end{figure*}

%% file: body/34-continuation.tex
\paragraph{Continuation}
\label{subsec:continuation}

A crucial question when filling the density grid from the elements is how large their region of influence should be. This choice impacts convergence significantly.

The parameters controlling the spatial influence of the elements are
$\rbfWidth$ and (to a lesser extent) $\rbfSlope$ in the sample's influence function given in \Cref{eq:material-densities:heaviside}. $\rbfWidth$ directly controls the width of the RBF.
%
%
If $\rbfWidth = 1$, then the material density given by $\densitySampleFunc(\currentPosition)$ tightly corresponds to the density of the physical element represented by the sample.

Setting $\rbfWidth = 1$ seems to be an obvious choice.
However, elements can only be attracted towards regions where they have a non zero contribution. Thus, using a tight RBF will impede the ability of the optimizer to pull elements towards regions of high compliance. In the worst case, some elements can end up floating in regions of low compliance while never be attracted where they would be most useful, as illustrated in \Cref{fig:continuation:none}.
By setting $\rbfWidth > 1$ we can artificially \emph{enlarge} the region an element spans. This unfortunately creates a situation where the RBFs are no longer representative of the physical elements.

To mitigate these effects while enabling a good convergence, we introduce a continuation method on the radius of influence multiplier $\rbfWidth$, as illustrated in \Cref{fig:continuation-density}. This is akin to multi-resolution optimization, where a coarse solution is produced and then refined. We start the gradient-based optimization with a high value of $\rbfWidth$ (3 in our implementation), and progressively decrease it every 30
iterations, multiplying $\rbfWidth$ by a factor of $0.9$ until it reaches a minimum value. We set the minimum value to $\rbfWidth=0.9$ to encourage small overlaps between adjacent elements -- which is necessary in order to print aggregate geometry. We treat $\rbfSlope$ -- that controls the smoothness of the density change -- in a similar fashion (see \Cref{alg:optimizer-main-loop}).

The effect of the continuation parameter $\rbfWidth$ can be observed in \Cref{fig:continuation}. Without it, some elements end up stranded away from the main structure.

\input{figs/continuation-density}

%% file: figs/continuation-density.tex
\begin{figure}[b]
	\centering
        \subfloat[element samples]{
	  \includegraphics[width=0.32\linewidth]{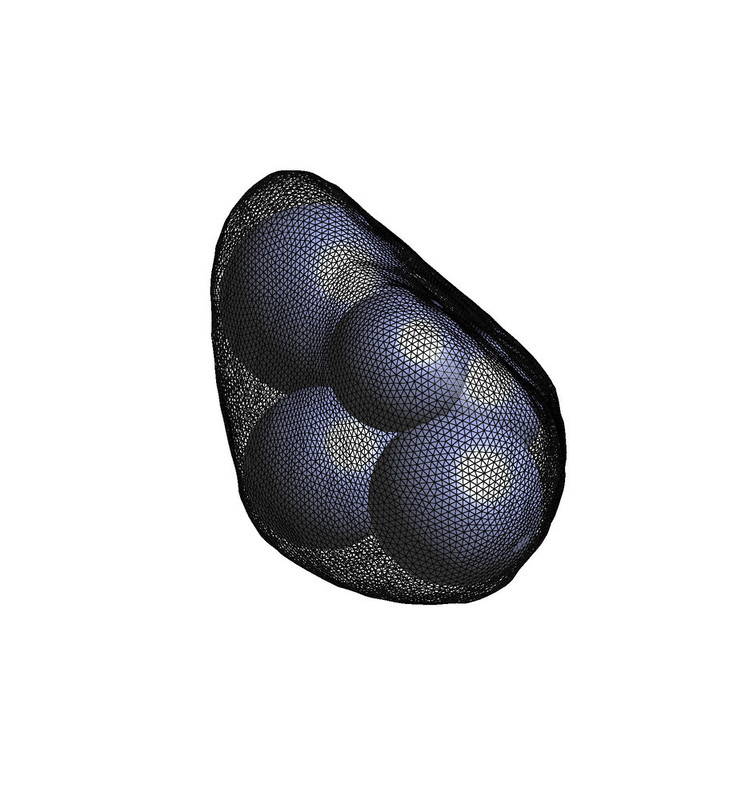}
        }
        \subfloat[density grid, $\rbfWidth = 3$]{
	  \includegraphics[width=0.32\linewidth]{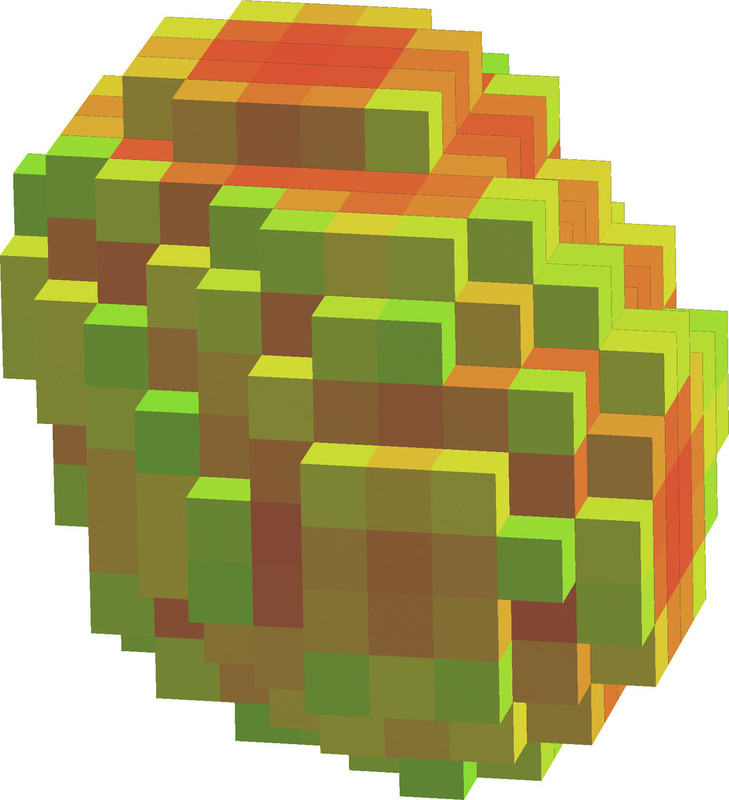}
        }
        \subfloat[density grid, $\rbfWidth = 1$]{
	  \includegraphics[width=0.32\linewidth]{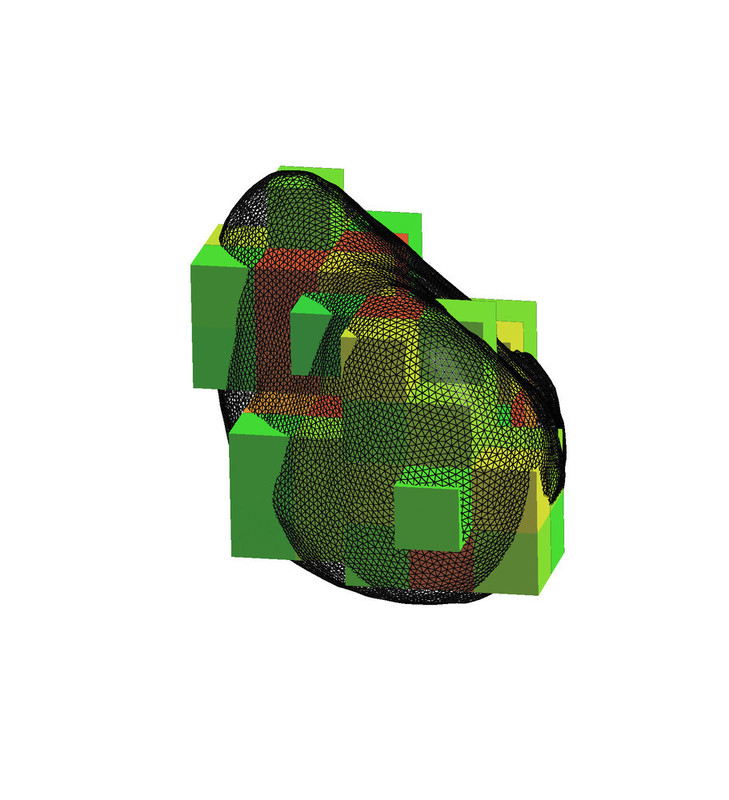}
        }
	\Caption{Effect of the continuation parameter $\rbfWidth$ on the material densities.}
        {%
		In the first iterations of the optimization, each element is set to occupy more physical space than it actually covers, to help the optimizer attract elements towards weak regions. Then, as we progressively reduce $\rbfWidth$ (by a factor of $0.9$ every $30$ iterations in our algorithm), the regular grid densities will match the actual physical object, up to the discretization error.
	}
	\label{fig:continuation-density}
\end{figure}

%% file: body/38-connect.tex
\paragraph{Connectivity}
\label{subsec:connectivity}

Even though the compliance optimization will naturally discourage small gaps between elements, the gaps might still occur due to the discretization.
Specifically, a gap sufficiently smaller than a grid cell can get filtered away and go unnoticed by the optimizer. To reduce this issue we embed in the optimization loop the following mechanism that explicitly encourages contacts.

The connectivity improvement algorithm works as follows: 1) detect pairs of samples from distinct elements that we would like to move towards each other, and define for each pair a target length so that elements overlap, 2) minimize the potential energy of a spring-mass system on the graph defined by those edges (we do 10 steps of BFGS every time at every continuation step, see \Cref{alg:optimizer-main-loop}).

We compute a graph $G = (\allSamples, \graphEdges, \graphWeight)$ following \Cref{alg:build-connectivity-graph}\iflabelexists{line:build-connectivity-graph}{~(called from line~\ref{line:build-connectivity-graph} in \Cref{alg:optimizer-main-loop})}{}, inspired by Kruskal's min-covering tree algorithm.
Then, we interpret $G$ as a spring-mass network with rest length $\graphWeight$, and minimize its potential energy via $10$ steps of a BGFS solver.

\input{figs/connectivity-algo}

%% file: figs/connectivity-algo.tex
\begin{algorithm}[htbp]
	\caption{\textsc{BuildConnectivityGraph}}
	\label{alg:build-connectivity-graph}

	\KwIn{Element parameters $\allDofs$, sample positions $\allSamplePos$ and their radii $\{\sampleRadius\}_\sample$.}
	\KwOut{Connectivity graph $G = (\allSamples, \graphEdges, \graphWeight)$.}

	$\graphEdges^{(0)} \gets \textsc{DelaunayTriangulation}(\allSamplePos).\texttt{edges}()$ \tcp*[l]{candidate connections}
	$\graphWeight_{ij} \gets \norm{\currentPosition_i - \currentPosition_j} \quad \forall ij \in \graphEdges$ \tcp*[l]{current distance between samples}
	$H \gets \textsc{UnionFind}()$ \tcp*[l]{union-find data structure for element ids}
	$\graphEdges \gets \emptyset$ \tcp*[l]{final set of edges for $G$}
	\ForEach{$ij \in \graphEdges^{(0)} \text{ by } \nearrow \graphWeight_{ij}$} {
		$\element_i \gets i.\texttt{elementId}() ; \element_j \gets j.\texttt{elementId}()$ \;
		\If{$\element_i \neq \element_j$}{
			\If{
				$H.\texttt{find}(\element_i) \neq H.\texttt{find}(\element_j)$ \Or
				$\size{\N_H(\element_i)} \leq 1$ \Or
				$\size{\N_H(\element_j)} \leq 1$
			}{
				$\graphEdges \gets \graphEdges \cup \{i,j\}$ \;
				$H.\texttt{merge}(\element_i, \element_j)$ \tcp*[l]{$\element_i$ and $\element_j$ are now neighbors in $H$}
			}
		}
	}
	\tcp{finalize: set target distances between samples}
	$\graphWeight_{ij} \gets 0.9 (\sampleRadiusSymbol_i + \sampleRadiusSymbol_j) \quad \forall ij \in \graphEdges$

	\Return{$G = (\allSamples, \graphEdges, \graphWeight)$}
\end{algorithm}

%% file: body/37-rotparam.tex
\subsection{Parameterization of 3D Rotations}
\label{subsec:rotparam}

We have to differentiate the sample positions $\allSamplePos$ with respect to the element parameters $\allDofs$, including 3D rotations.
We parameterize them using exponential maps, and in particular the formulation given by~\cite{Grassia:1998:PPO} which computes the exponential map from $\lieAlgebra(3)$ to $\rotationGroup(3)$ via an intermediate quaternion representation.
The authors provide a \cLang{} implementation for computing the partial derivatives of the rotation matrix with respect to the exponential map vector in $\lieAlgebra(3)$.
%
Since the exponential maps can become ill-conditioned if the rotation is too high, we perform a check every 30 iterations of our algorithm, and we reparameterize rotations that have become too large (angle $\geq \pi$). More specifically, we set $\elementLinear \gets \elementRotation\elementLinear$ and $\elementRotation \gets \identity$.

%% file: body/40-results.tex
\input{body/44-numelement-fig}
\input{body/46-numiteration-fig}
\input{body/43-input-fig}
\input{body/41-result-fig}

\input{figs/results-noodles}
\input{body/42-sword-chair-fig}

\input{figs/results-bookend}
\input{figs/results-teapot}
\input{figs/results-chairs}

\section{Results}
\label{sec:results}

In this section we first discuss the influence of various parameters, provide timings and statistics on our results, and then show a number of 3D printed results (on a ZCorp 450 color powder printer). We also provide renderings of additional results. The accompanying video shows animations of the optimization process as well as rotating views of results.
The inputs used throughout this section are visible in \Cref{fig:teaser} and \Cref{fig:input}.

\paragraph{Analysis}

The main parameter the user has to choose is the number of elements, and thus, indirectly, the solid volume percentage. \Cref{fig:numelement} shows the influence on the result for the chair case. As can be seen, the optimizer successfully arranges the elements in rigid connected structures, even on the result with the smallest number of elements. Of course, more elements affords for a more sturdy structure as more material is available.

\Cref{fig:numiteration} visualizes the behavior of the optimization algorithm as it iterates. Animations are available in the accompanying video.

\Cref{fig:continuation} shows the benefits of our continuation and connectivity methods (\Cref{subsec:continuation,subsec:connectivity}).
In particular, a few stranded elements are visible when continuation is disabled, while the connectivity step encourages small overlaps between adjacent elements.

Timings are reported in \Cref{tab:timings}. All results were obtained on an Intel\textsuperscript{\textregistered} Core\texttrademark{} i7-5930K @ 3.50GHz, 64 GB RAM. Note that the first iterations are slower because RBFs $\densitySampleFunc(\currentPosition)$ have a larger support due to our continuation scheme (our evaluation procedure uses a single-thread on the CPU). As $\rbfWidth$ decreases, the iterations become faster, and are dominated by the cost of solving \Cref{eq:fem-equilibrium}.

\paragraph{Synthesized shapes}

As synthesis is fully automatic, it is very simple for the user to produce a variety of results by combining different elements and structural problems.
We synthesized and 3D printed several such examples.

The chair is a common daily object with different geometry components, including thin legs and planar seats and backs. Our method automatically assembles a chair from its structural definition, out of different elements including mixture of fruits (\Cref{fig:result:masterchairs}, top),
wood sticks (\Cref{fig:result:woodstick_chair}),
pebbles (\Cref{fig:result:pebble_chair}),
long flexible noodles (\Cref{fig:results:noodles}, \Cref{fig:result:masterchairs} bottom) and even swords (\Cref{fig:sword-chair}). This later case is challenging as it uses a large number of sharp, elongated swords. Our method succeeds in maintaining a sound structure where the elements remain easily identifiable.

We also synthesize and 3D print table structures from
elongated sticks (\Cref{fig:result:woodstick_bridge}) and
pebbles (\Cref{fig:result:pebble_bridge}, \Cref{fig:results:teapot}), and created a bookend made of fruits, rendered in \Cref{fig:results:bookend}.

All these results were printed or rendered without any change after optimization. However, it is worth mentioning that since the output is made of elements, it would be simple to create a tool allowing the user to select, move, scale, delete or add some elements.


\input{figs/timings}


%% file: body/44-numelement-fig.tex
\begin{figure}[t]
  \centering
  \subfloat[100 elements]{
    \includegraphics[width=0.33\linewidth]{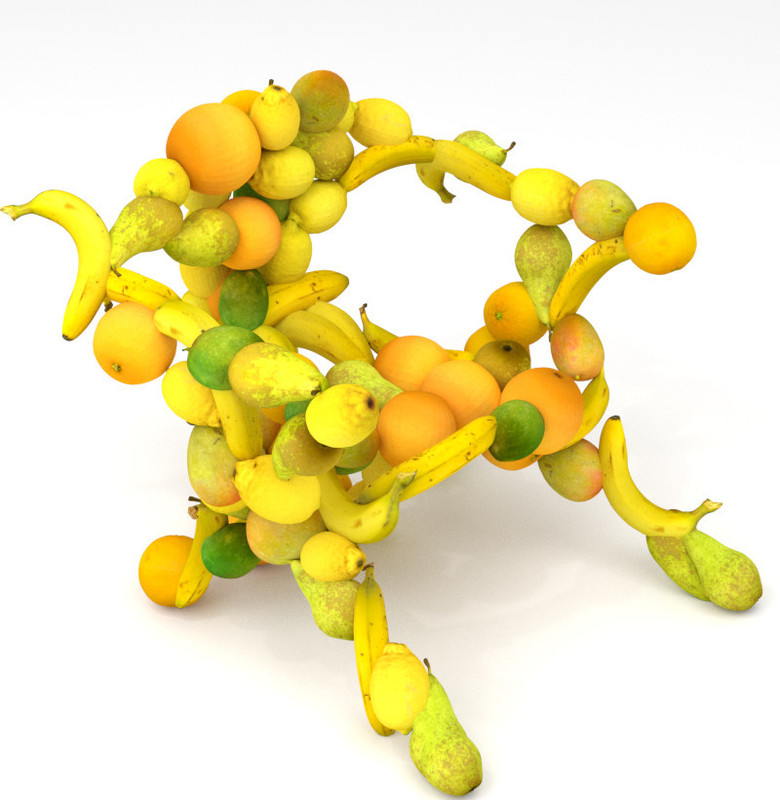}
  }
  \subfloat[150 elements]{
    \includegraphics[width=0.33\linewidth]{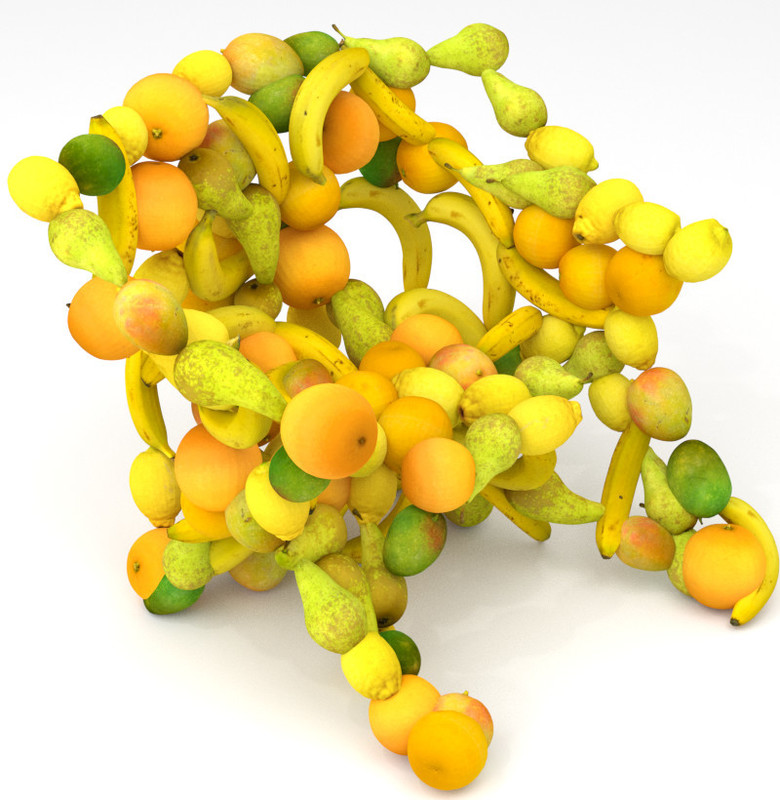}
  }
  \subfloat[200 elements]{
    \includegraphics[width=0.33\linewidth]{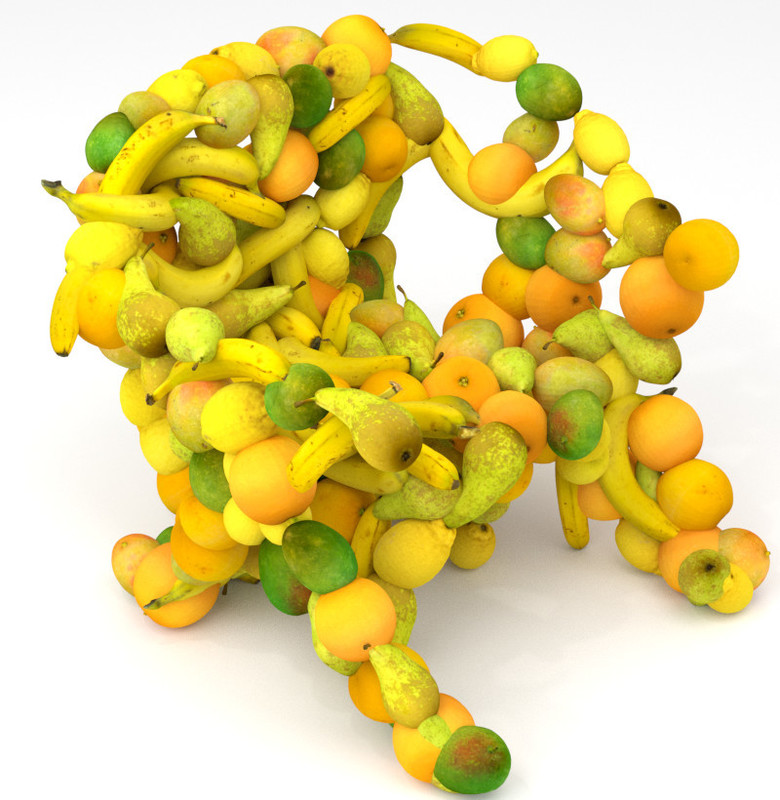}
  }
  \vspace*{-2mm}
  \Caption{Comparison of results with different number of elements under the same loading scenario (chair case).}
  {%
    Ten different elements (shown in \Cref{fig:input}) are used to produce results with respectively $100$, $150$, and $200$ elements.
  }
  \label{fig:numelement}
\end{figure}

%% file: body/46-numiteration-fig.tex
\begin{figure}[t]
  \centering
  \subfloat[iteration 0]{
	\includegraphics[width=0.33\linewidth]{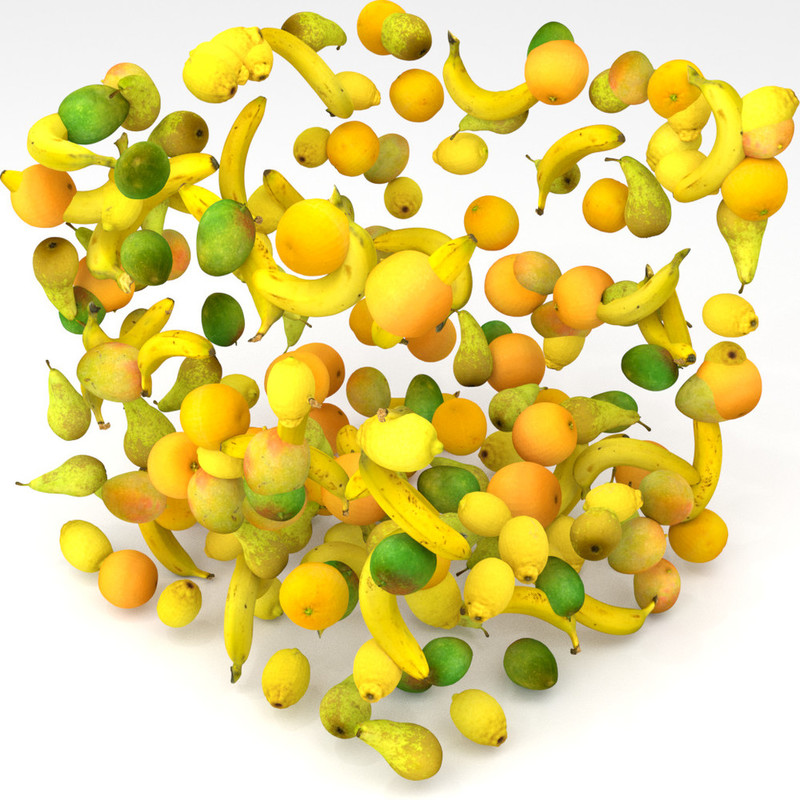}
  }
  \subfloat[iteration 60]{
	\includegraphics[width=0.33\linewidth]{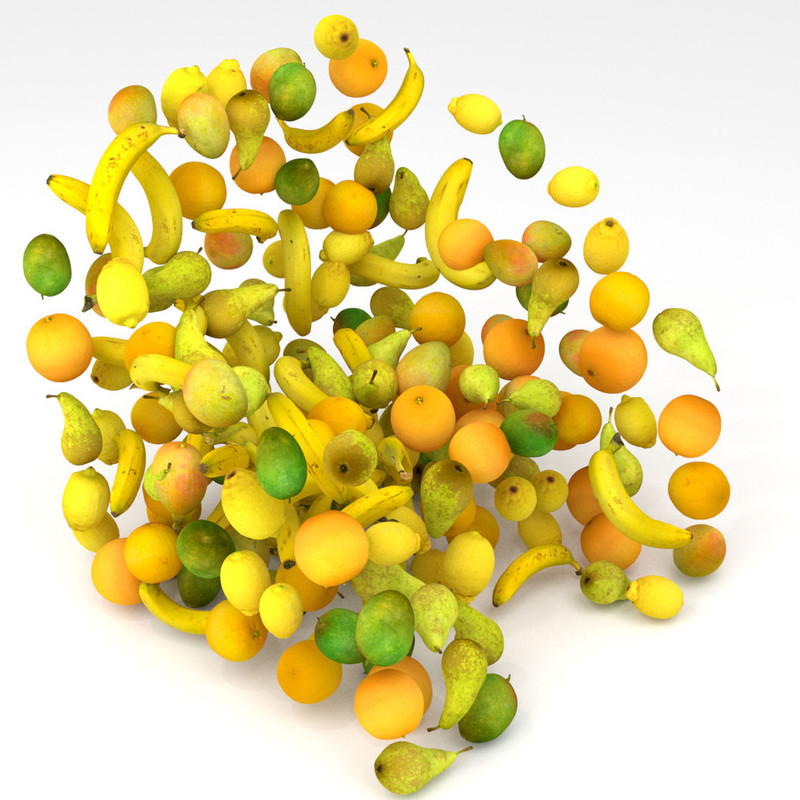}
  }
  \subfloat[iteration 500]{
	\includegraphics[width=0.33\linewidth]{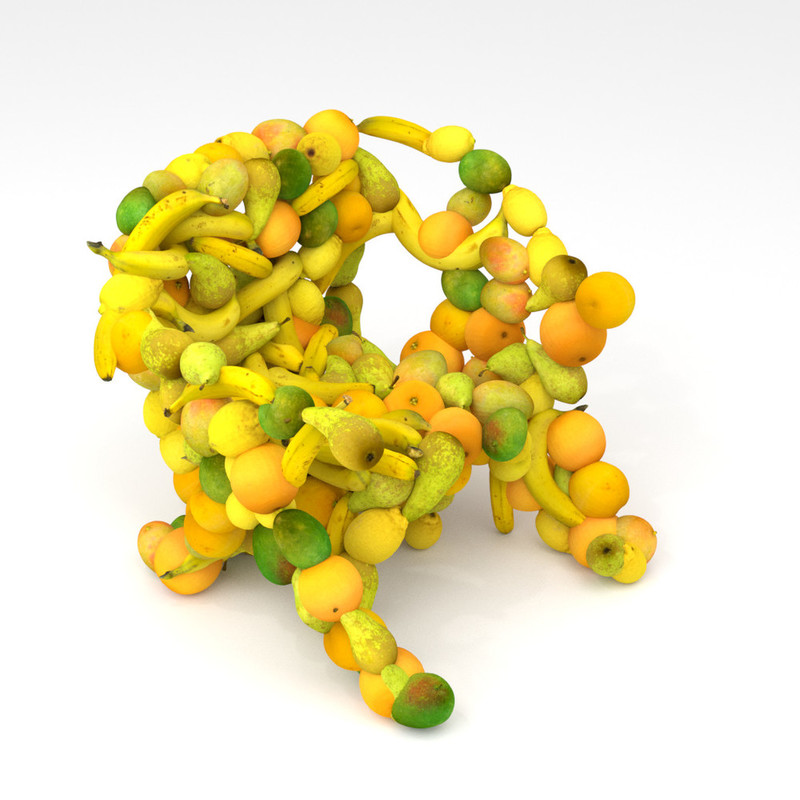}
  }
  \Caption{The effect of iterations.}
  {%
    Initially the elements tend to be floating and detached from one another. With more iterations they gradually form a stronger structure.
  }
  \label{fig:numiteration}
  \vspace*{-5mm}
\end{figure}

%% file: body/43-input-fig.tex
\begin{figure}[t]
	\centering
        \subfloat[bananas]{
	  \includegraphics[width=0.09\linewidth]{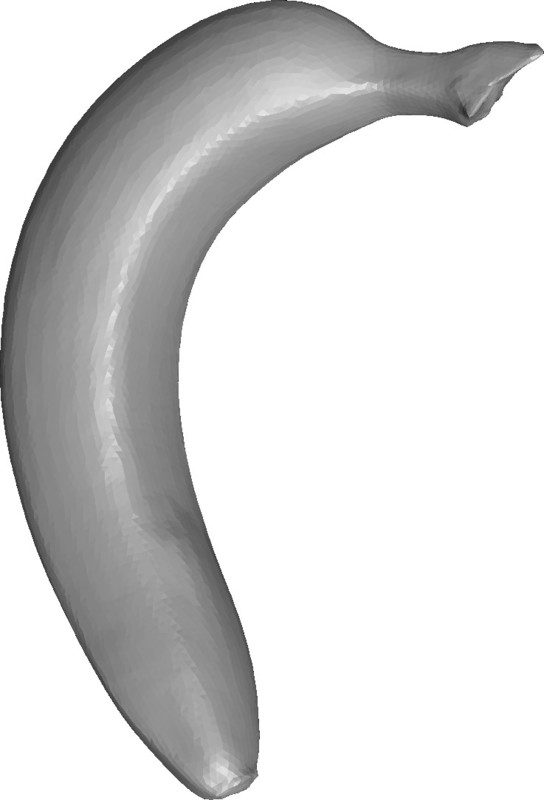}
	  \includegraphics[width=0.09\linewidth]{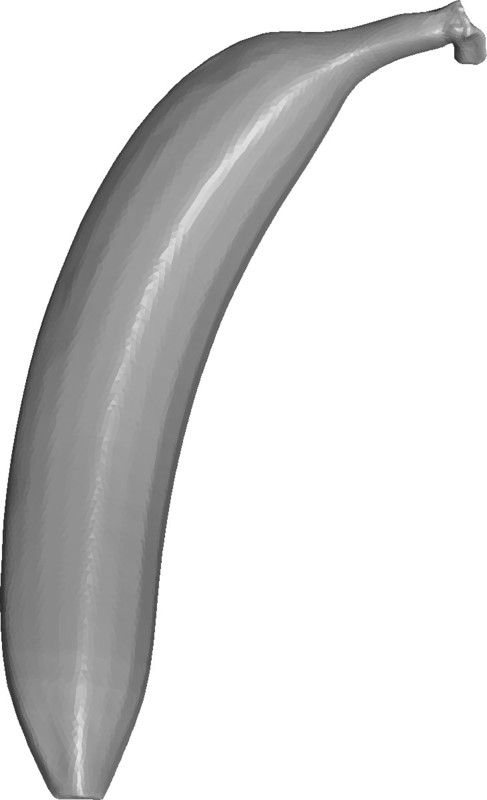}
        }
        \subfloat[lemons]{
	  \includegraphics[width=0.09\linewidth]{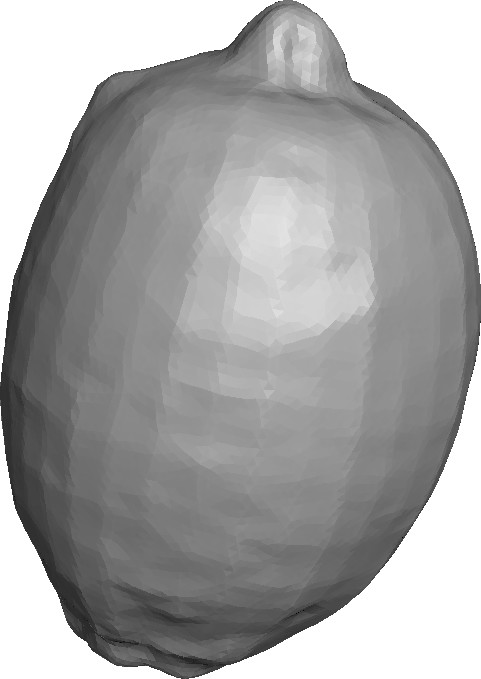}
	  \includegraphics[width=0.09\linewidth]{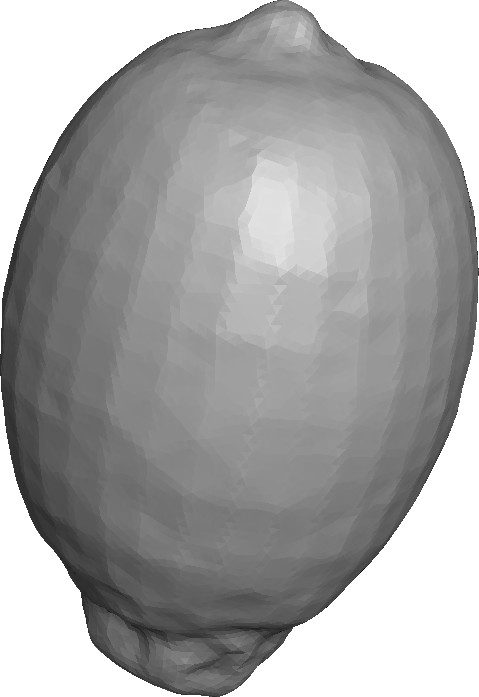}
        }
        \subfloat[mangos]{
	  \includegraphics[width=0.09\linewidth]{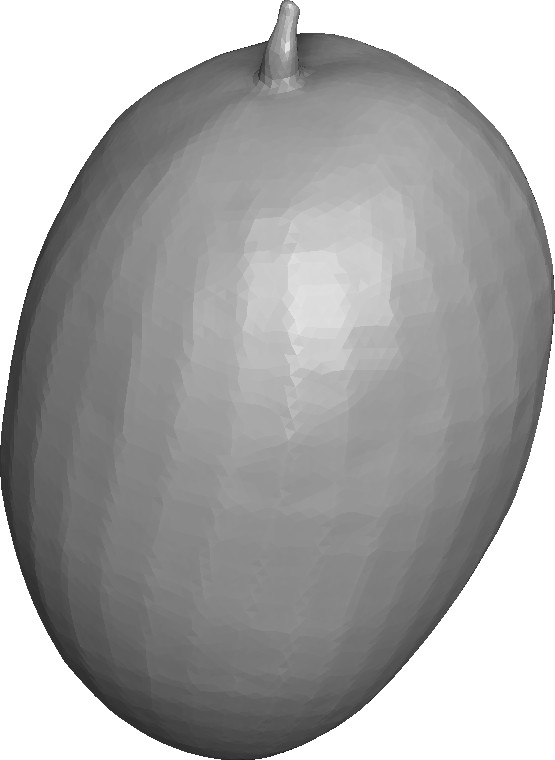}
	  \includegraphics[width=0.09\linewidth]{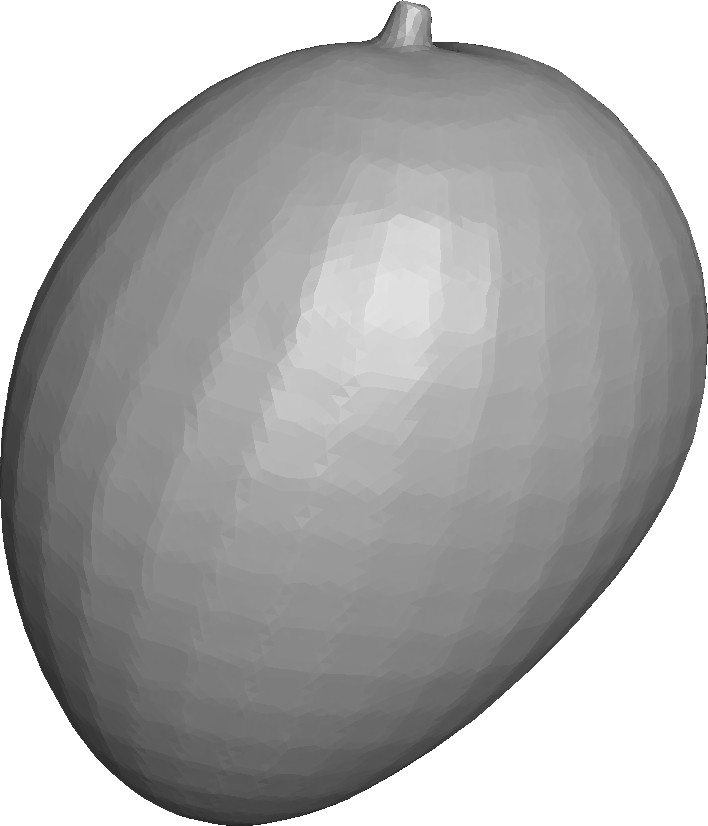}
        }
        \subfloat[oranges]{
	  \includegraphics[width=0.09\linewidth]{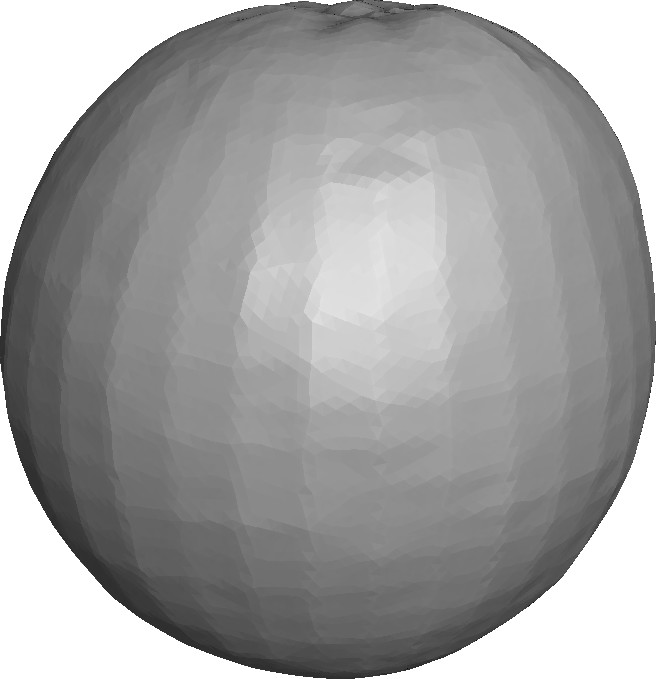}
	  \includegraphics[width=0.09\linewidth]{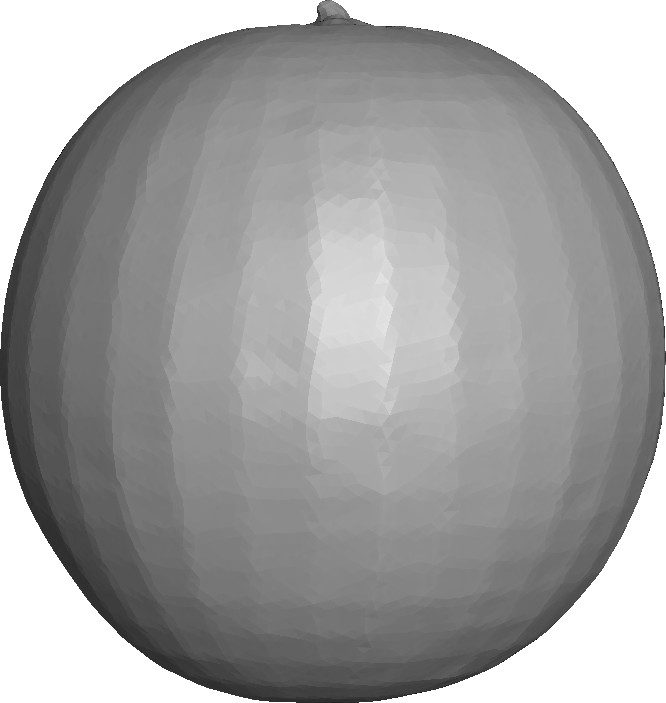}
        }
        \subfloat[pears]{
	  \includegraphics[width=0.09\linewidth]{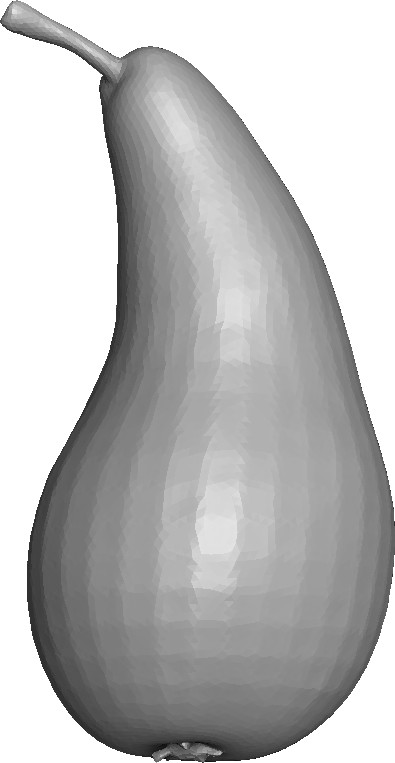}
	  \includegraphics[width=0.09\linewidth]{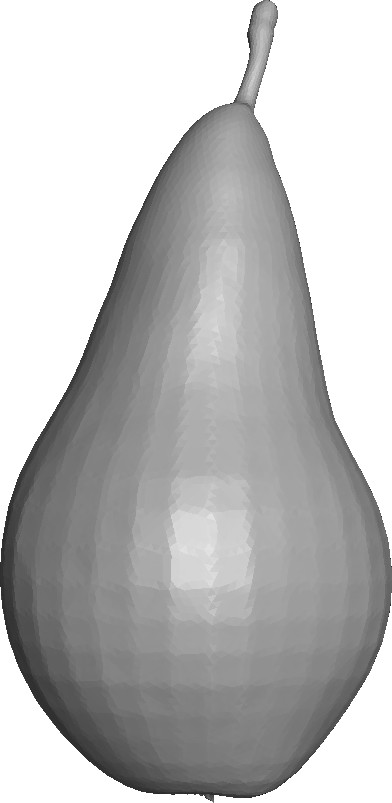}
        }

        \subfloat[swords]{
	  \includegraphics[width=0.2\linewidth]{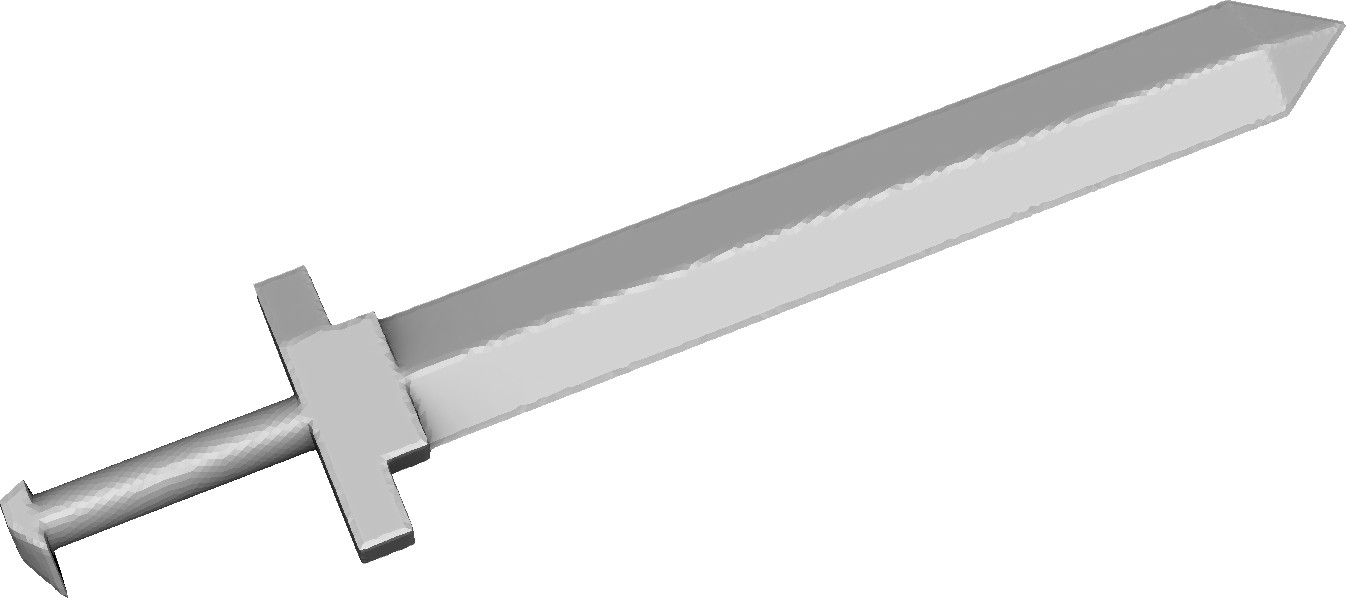}
	  \includegraphics[width=0.2\linewidth]{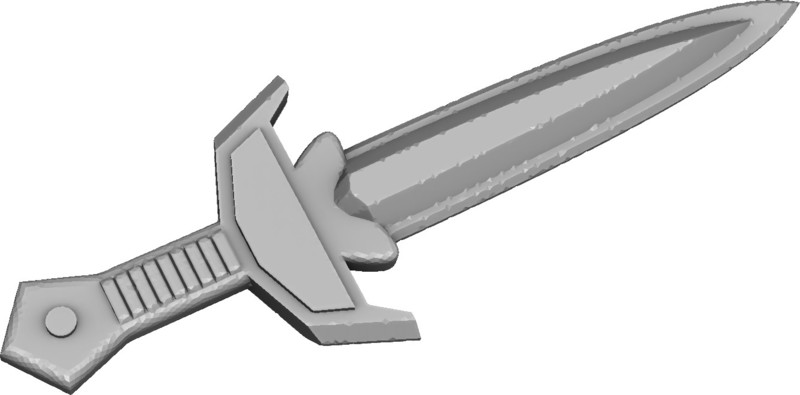}
	  \includegraphics[width=0.2\linewidth]{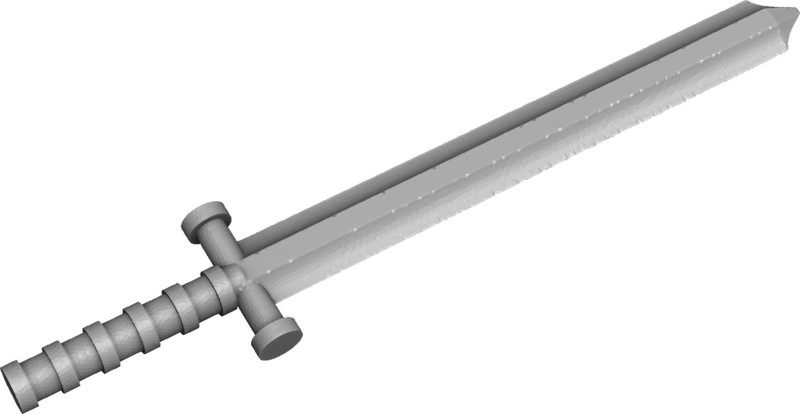}
        }
        \subfloat[woodstick]{
	  \includegraphics[width=0.2\linewidth]{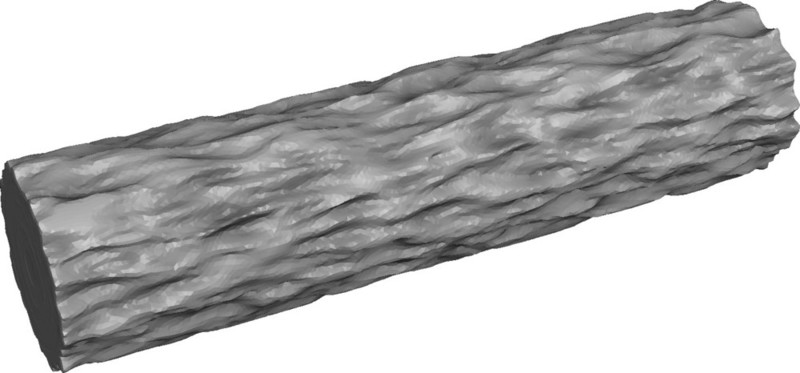}
        }
        \subfloat[helix]{
	  \includegraphics[width=0.2\linewidth]{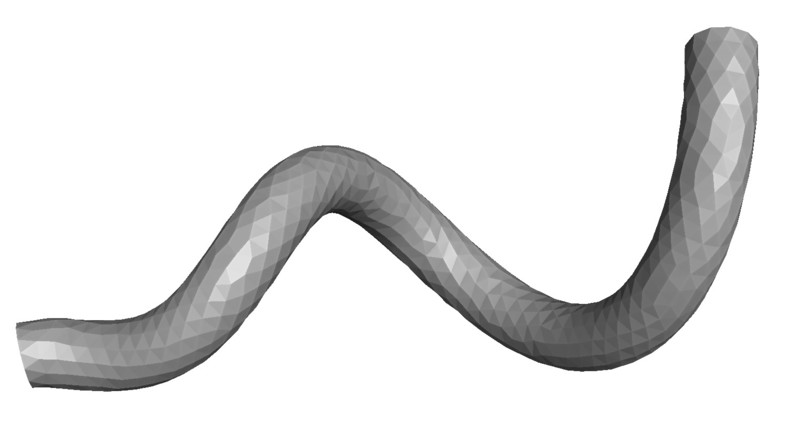}
        }

        \subfloat[table problem domain]{
	  \includegraphics[height=3cm]{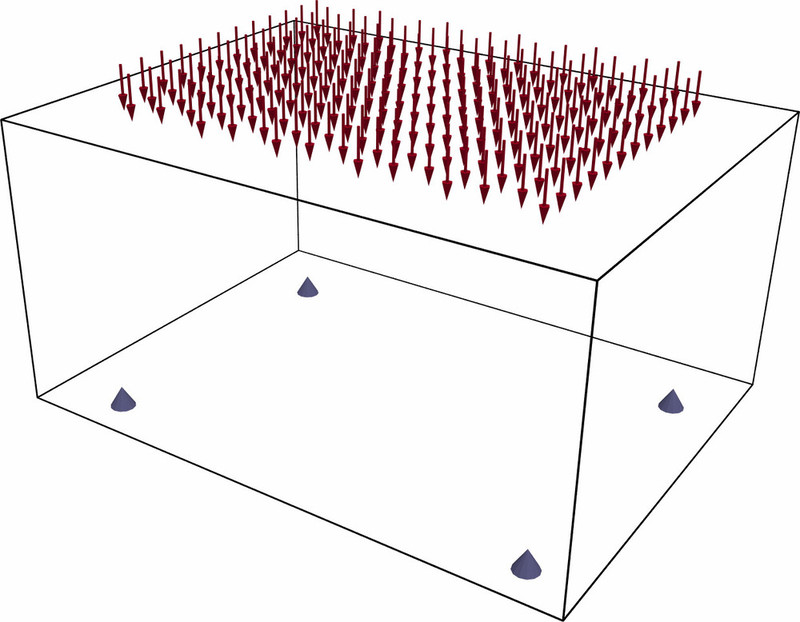}
        }
        \subfloat[bookend problem domain]{
          \label{fig:bookend_problem_domain}
      \includegraphics[height=3cm]{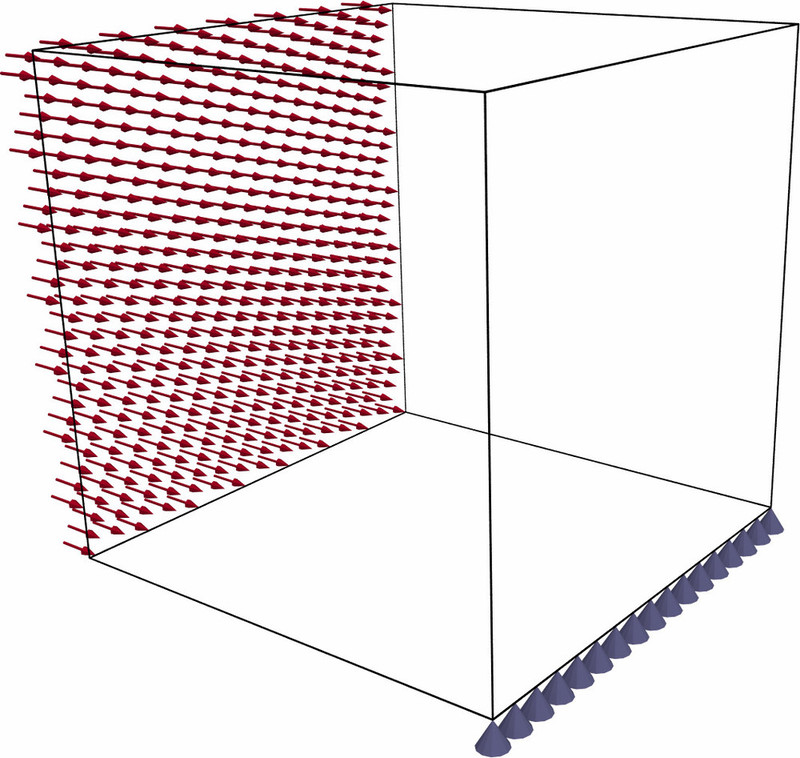}
        }
  \vspace*{-3mm}
  	\caption{Input elements and structural problem definitions.}
	\label{fig:input}
\end{figure}

%% file: body/41-result-fig.tex
\begin{figure}[tb]
  \centering

  \subfloat[woodstick chair]{
    \label{fig:result:woodstick_chair}
    \includegraphics[width=0.48\linewidth]{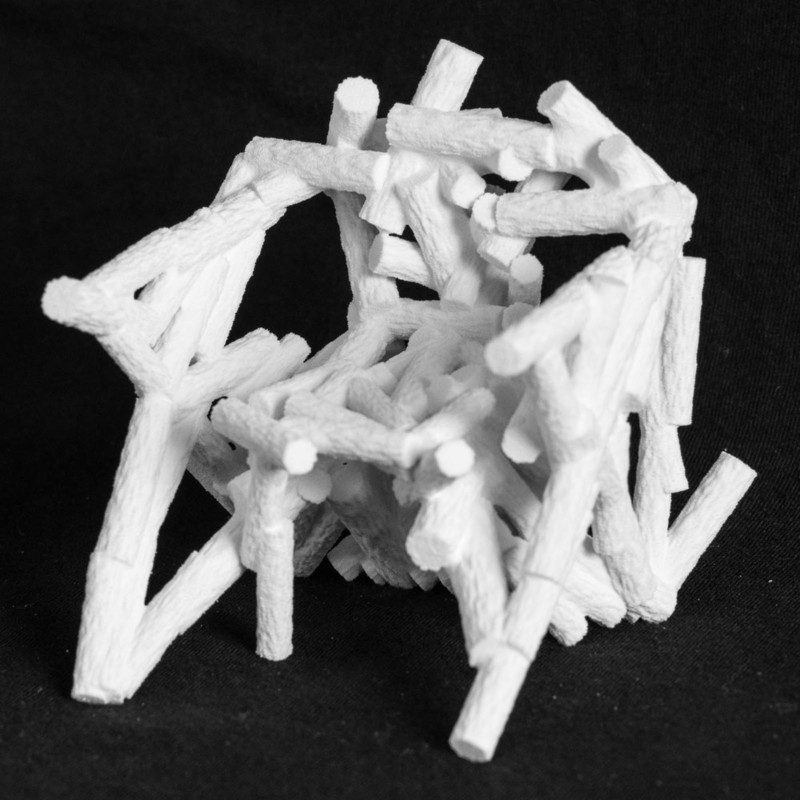}
    \includegraphics[width=0.48\linewidth]{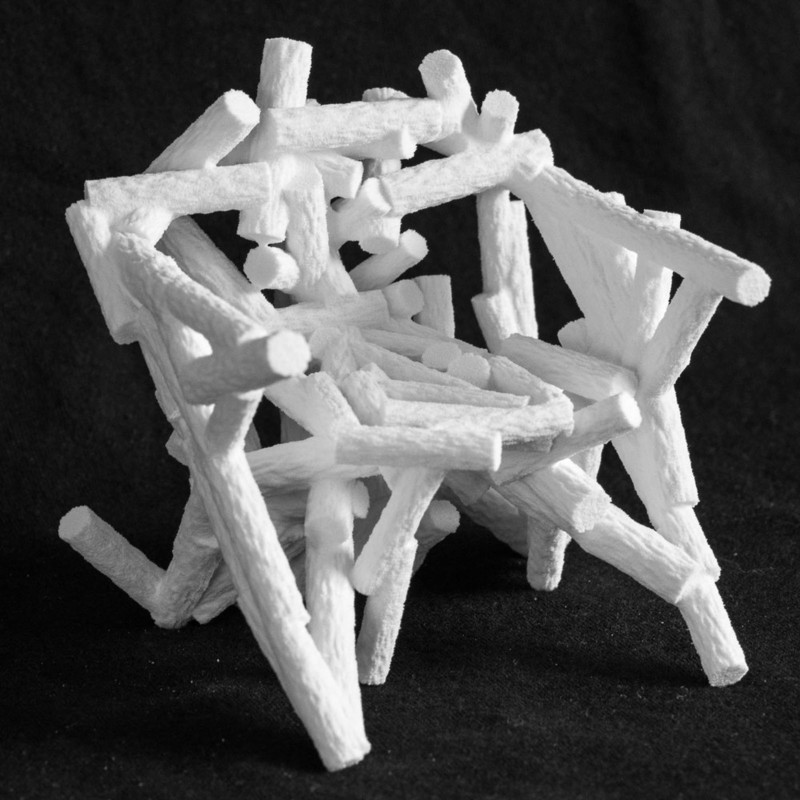}
  }

  \subfloat[woodstick table]{
    \label{fig:result:woodstick_bridge}
    \includegraphics[width=0.48\linewidth]{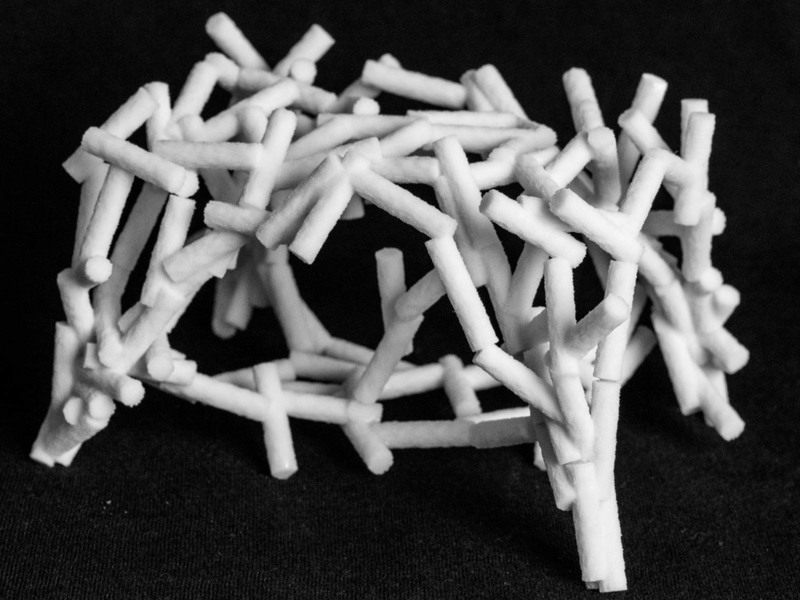}
    \includegraphics[width=0.48\linewidth]{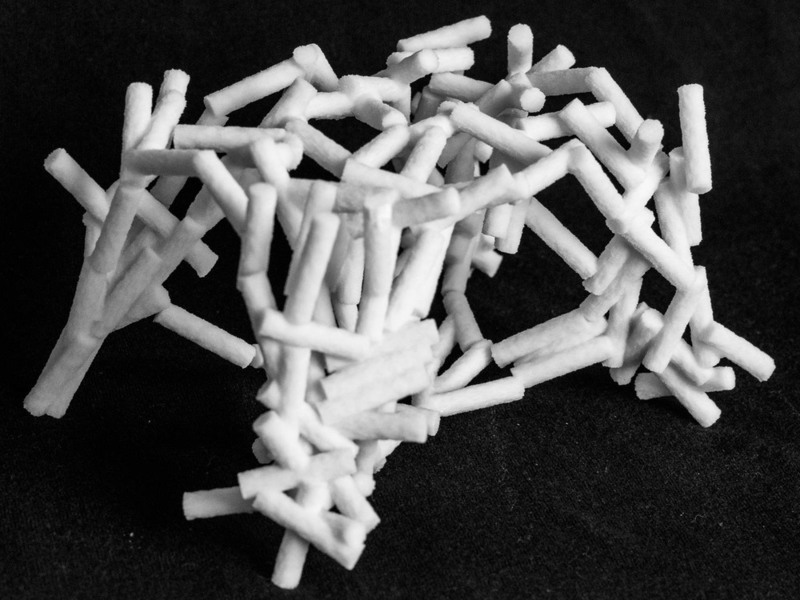}
  }

  \subfloat[pebble table]{
    \label{fig:result:pebble_bridge}
    \includegraphics[width=0.48\linewidth]{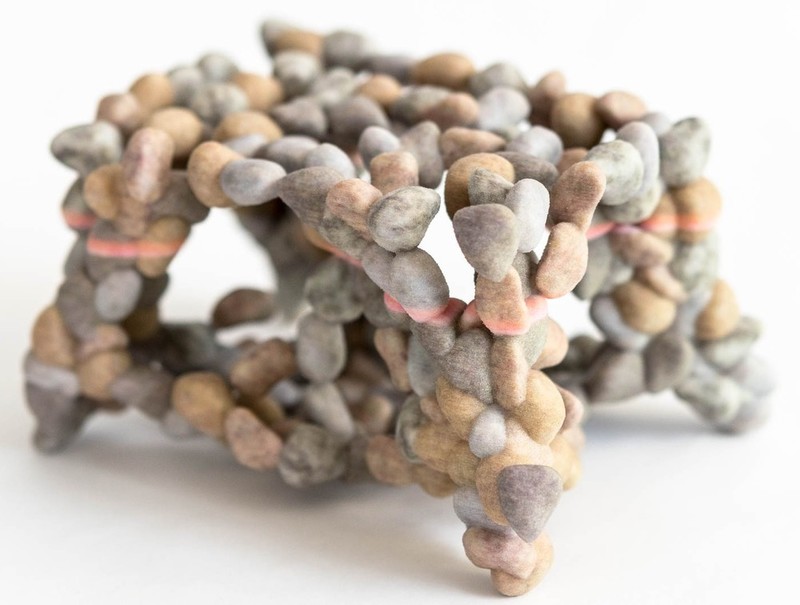}
    \includegraphics[width=0.48\linewidth]{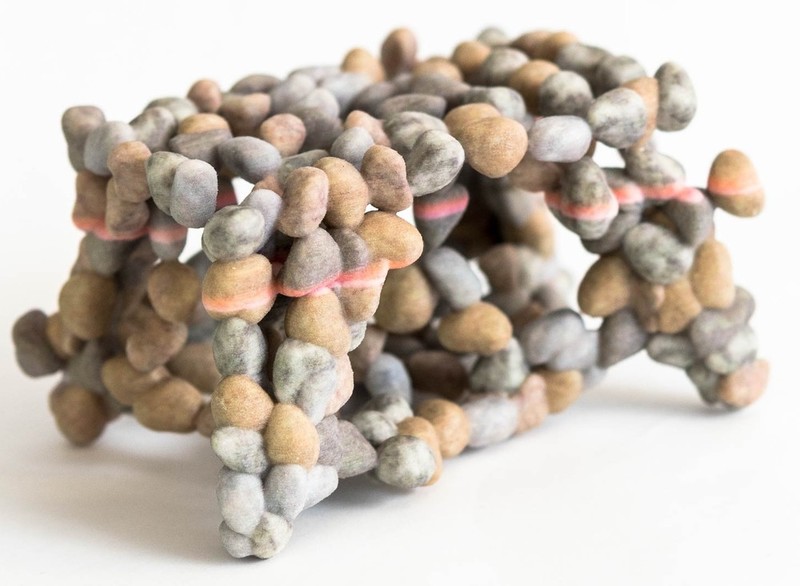}
  }
\vspace*{-2mm}
  \Caption{3D printed results.}
  {%
All results are manufactured on a powder printer, using colors for
\subref{fig:result:pebble_bridge} and without for \subref{fig:result:woodstick_chair} and \subref{fig:result:woodstick_bridge}.
  }
  \label{fig:result}
\end{figure}

%% file: figs/results-noodles.tex
\begin{figure}[tbh]
	\centering

	\subfloat[small deformation, $\abs{\dofAngularPos_{\sample_i}} \leq 0.15 \pi$]{
		\includegraphics[width=0.48\linewidth]{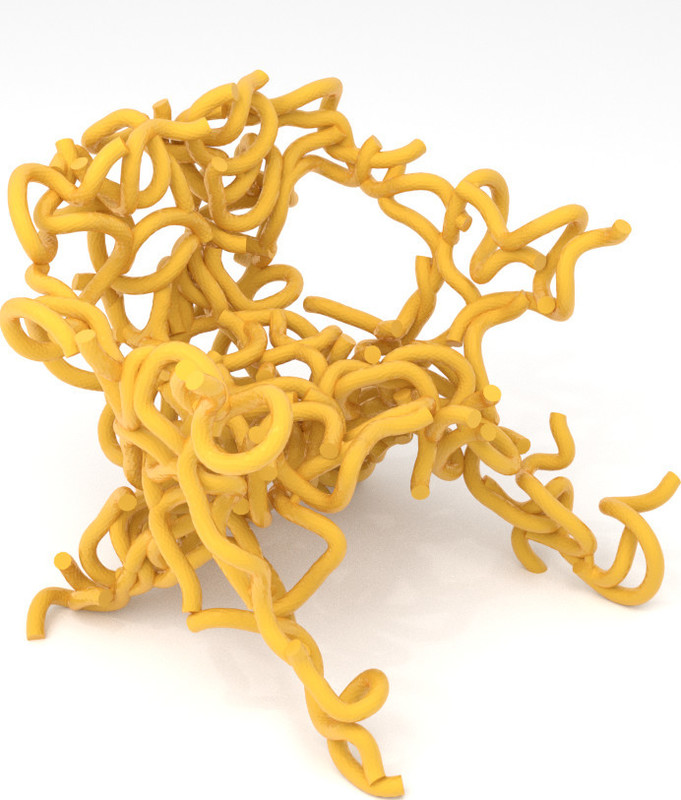}
	}
	\subfloat[larger deformation, $\abs{\dofAngularPos_{\sample_i}} \leq 0.3 \pi$]{
		\includegraphics[width=0.48\linewidth]{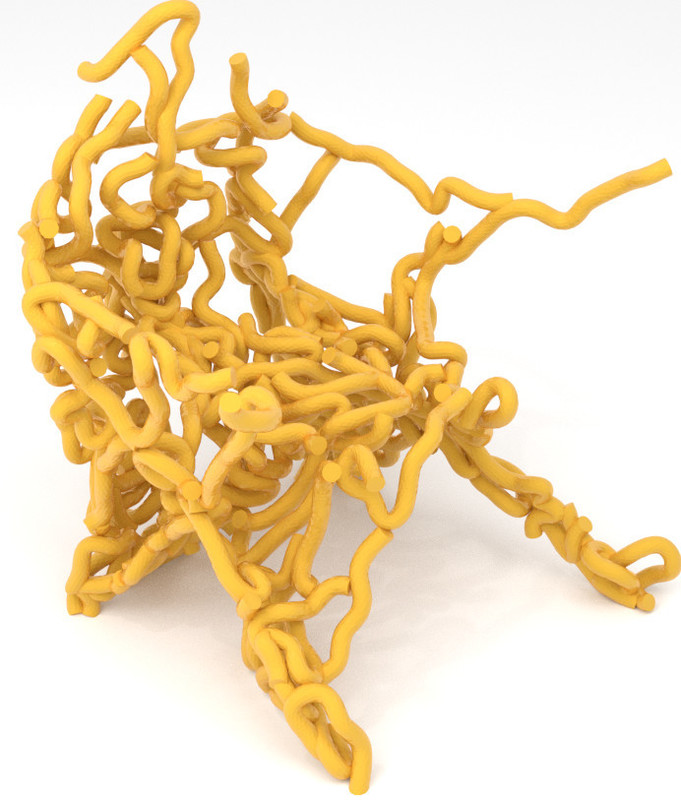}
	}
	\vspace*{-3mm}
	\Caption{A chair made of flexible noodles.}
        {%
          Different results can be achieved by allowing different amounts of deformation.
        }
	\label{fig:results:noodles}
	\vspace*{-2mm}
\end{figure}

%% file: body/42-sword-chair-fig.tex
\begin{figure}[tb]
  \centering
%
    \begin{minipage}{\linewidth}
      \centering
      \includegraphics[width=0.48\linewidth]{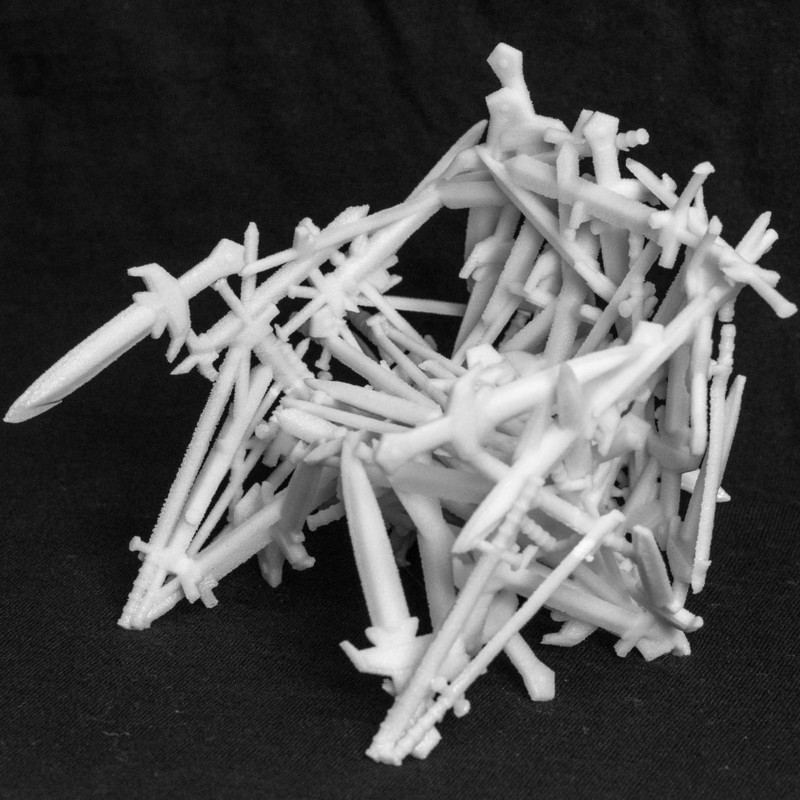}
      \includegraphics[width=0.48\linewidth]{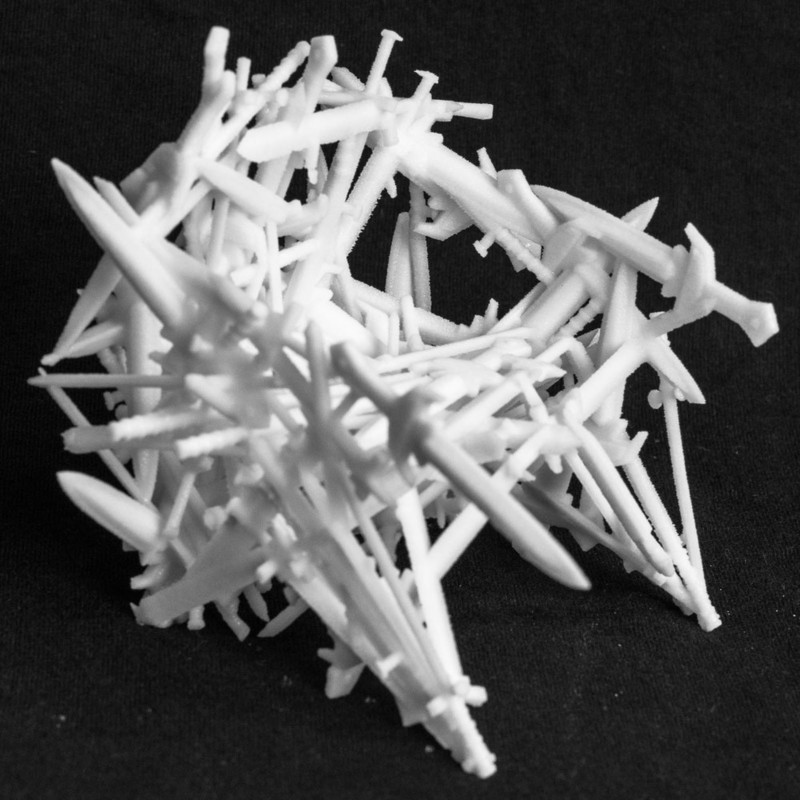}

      \includegraphics[width=0.48\linewidth]{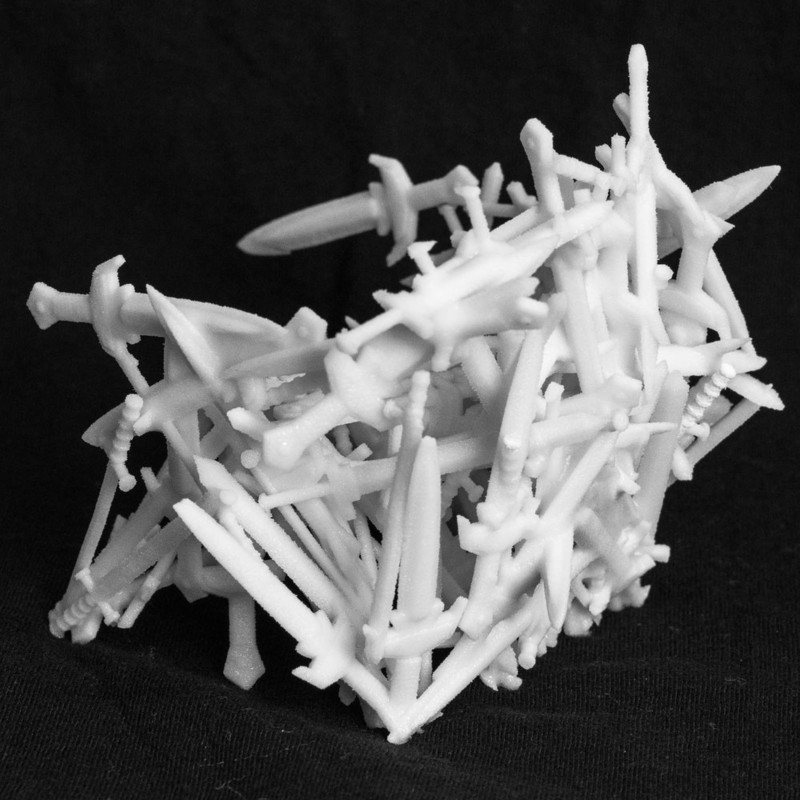}
      \includegraphics[width=0.48\linewidth]{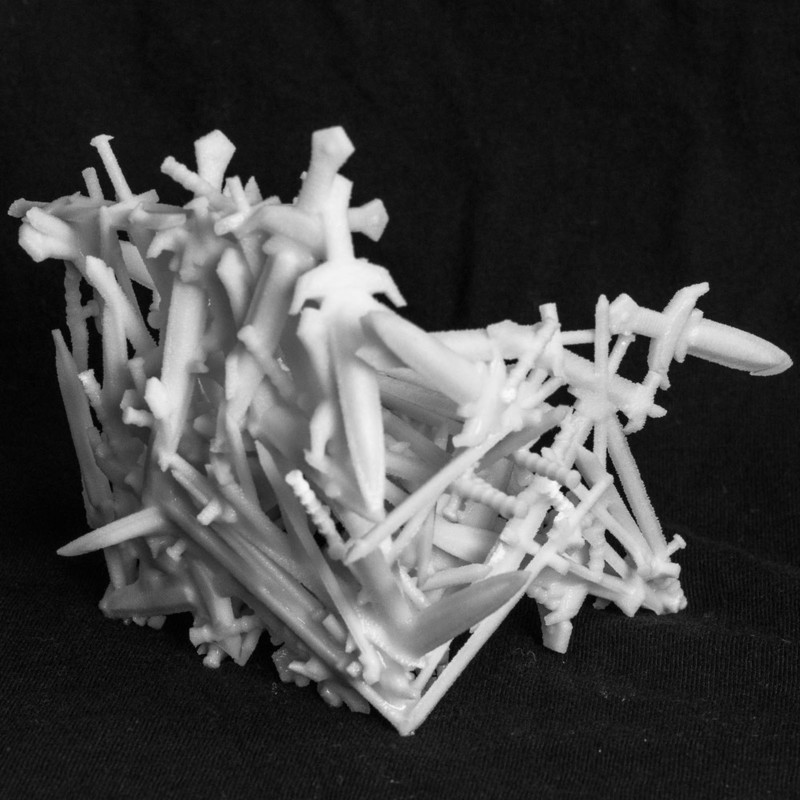}
    \end{minipage}
%
%
  \vspace*{-3mm}
  \Caption{A 3D printed iron throne made of swords.}{Cushion advised.}
  \label{fig:sword-chair}
\end{figure}

%% file: figs/results-bookend.tex
\begin{figure}[tbh]
	\centering

	\includegraphics[width=0.8\linewidth]{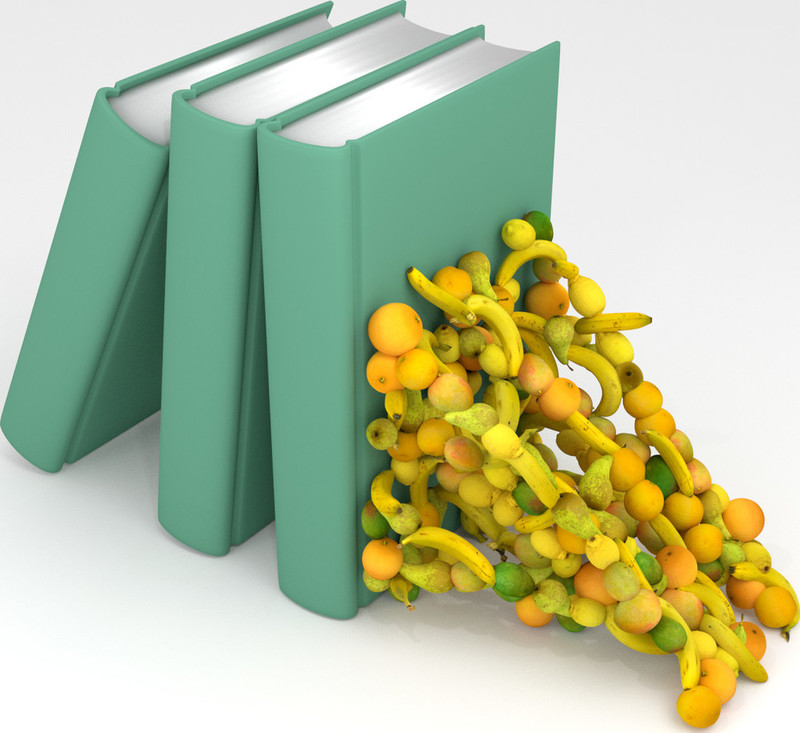}

	\Caption{Bookend made of fruits.}
        {%
		The loading scenario is shown in \Cref{fig:bookend_problem_domain}.
	}
	\label{fig:results:bookend}
\end{figure}

%% file: figs/results-teapot.tex
\begin{figure}[tbh]
	\centering

	\includegraphics[width=0.7\linewidth]{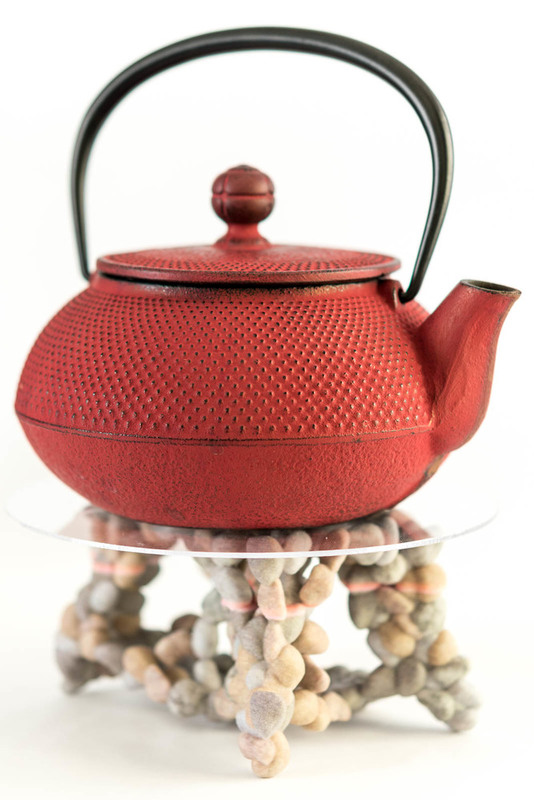}

	\caption{
		A 3D-printed table made of pebbles supporting a \SI{1.2}{\kilogram} teapot.
	}
	\label{fig:results:teapot}
\end{figure}

%% file: figs/results-chairs.tex
\begin{figure*}
    \includegraphics[width=0.24\linewidth]{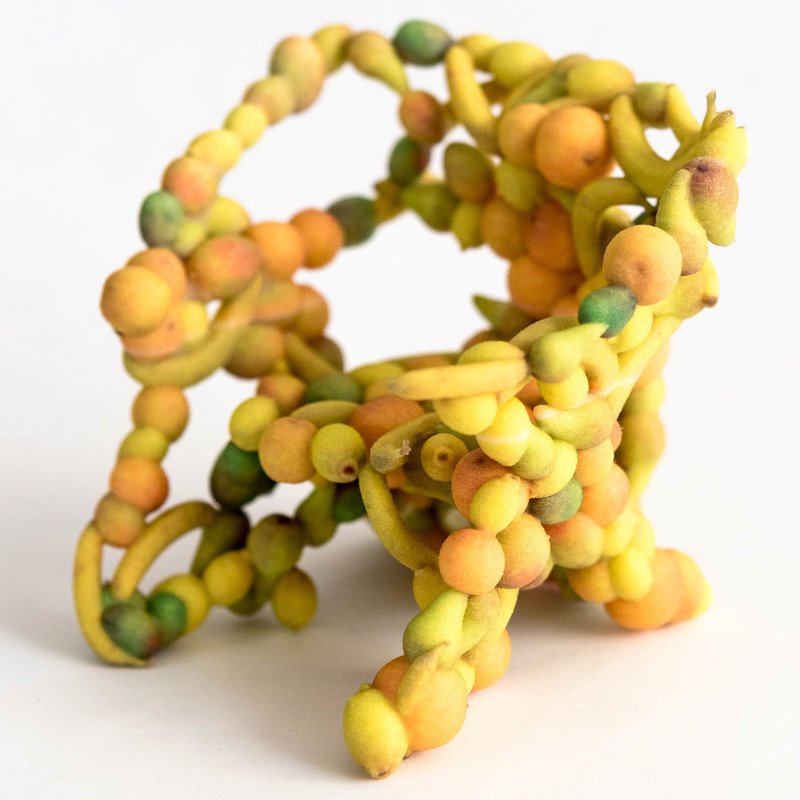}
    \includegraphics[width=0.24\linewidth]{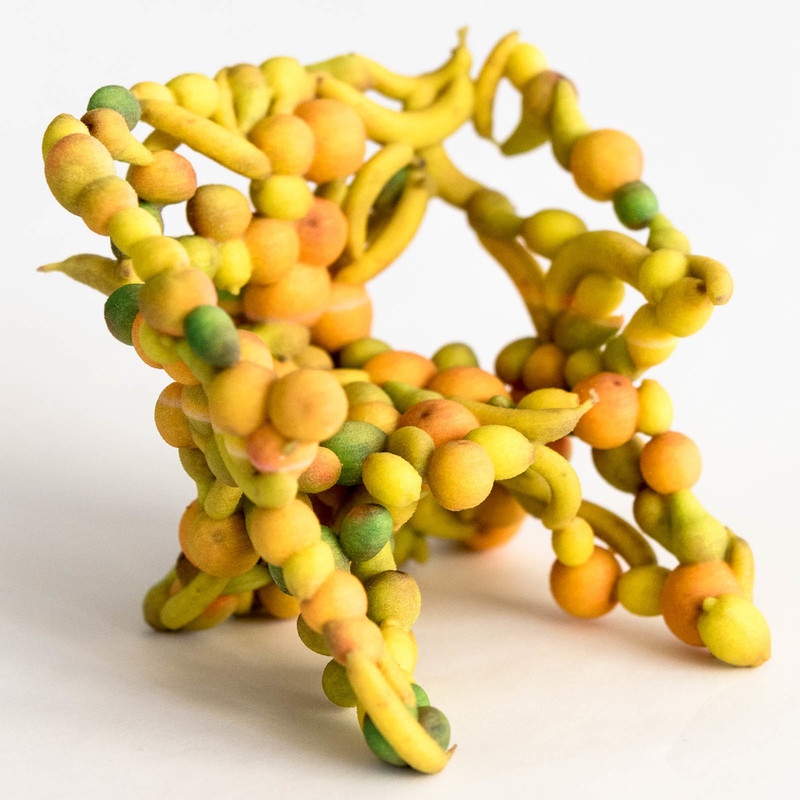}
    \includegraphics[width=0.24\linewidth]{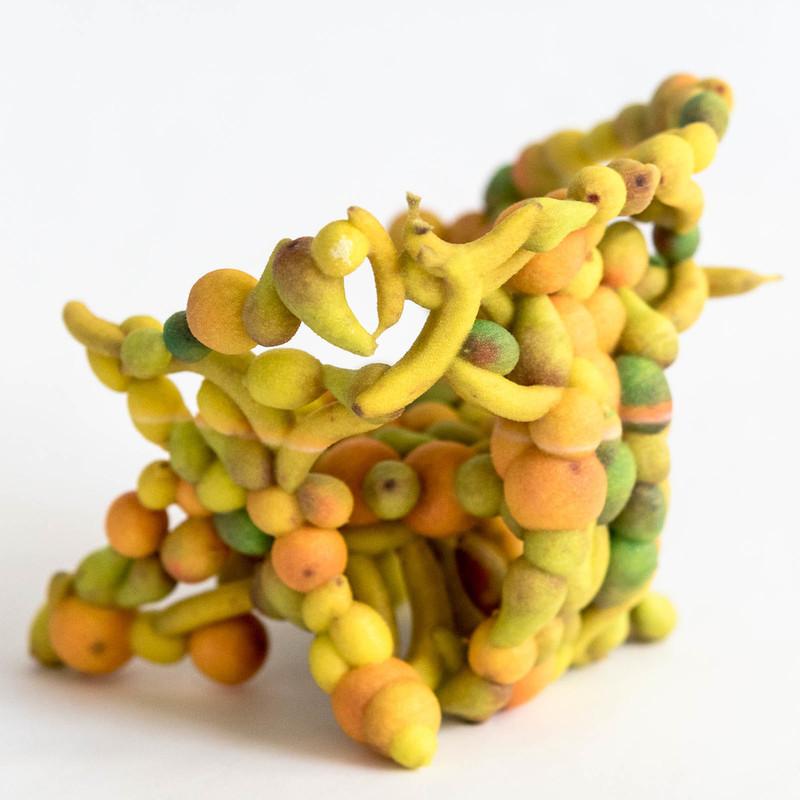}
    \includegraphics[width=0.24\linewidth]{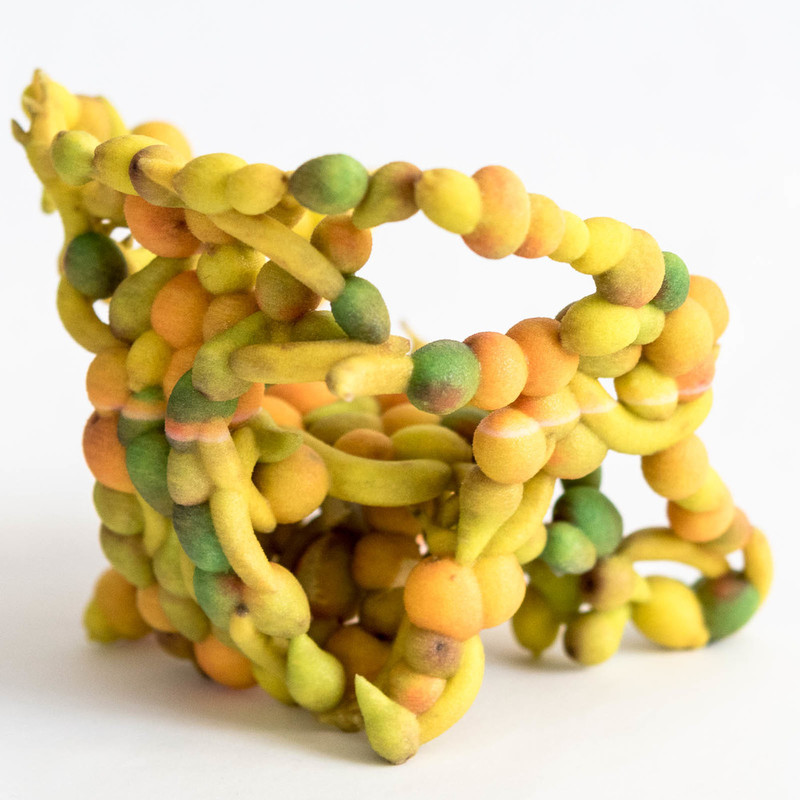}\\
    \includegraphics[width=0.24\linewidth]{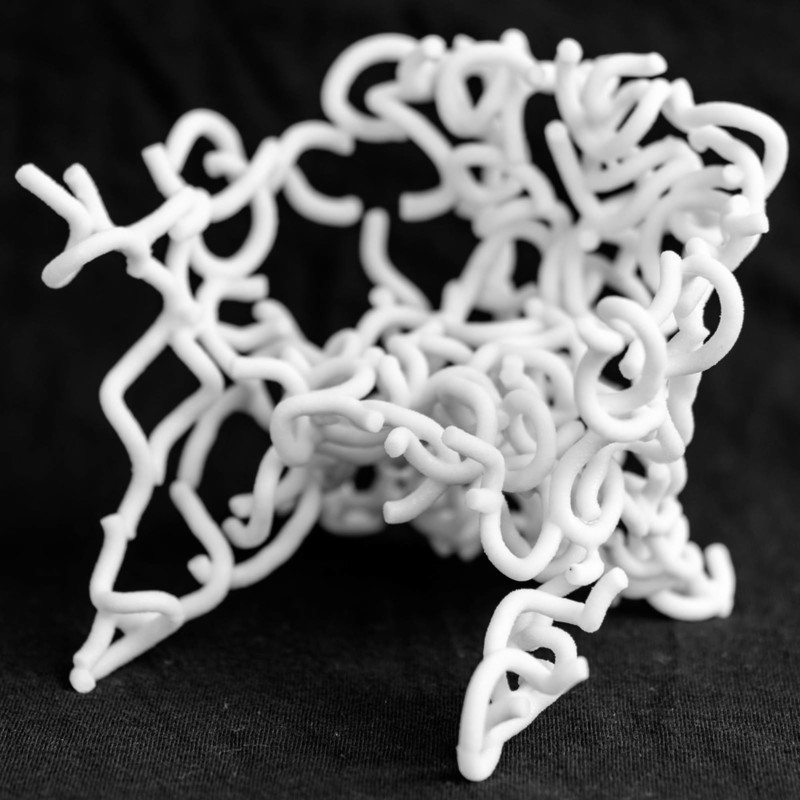}
    \includegraphics[width=0.24\linewidth]{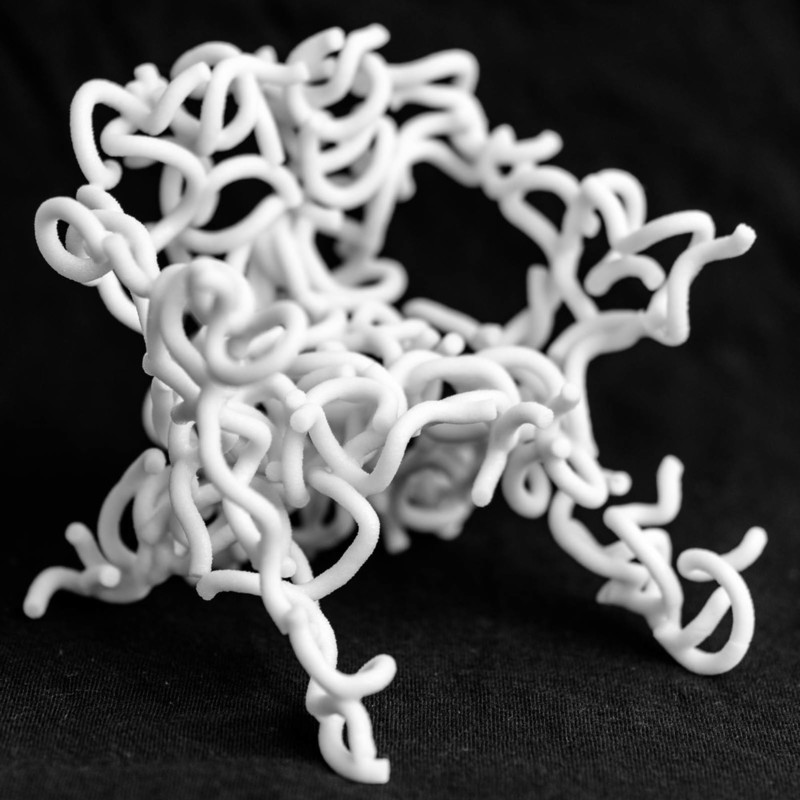}
    \includegraphics[width=0.24\linewidth]{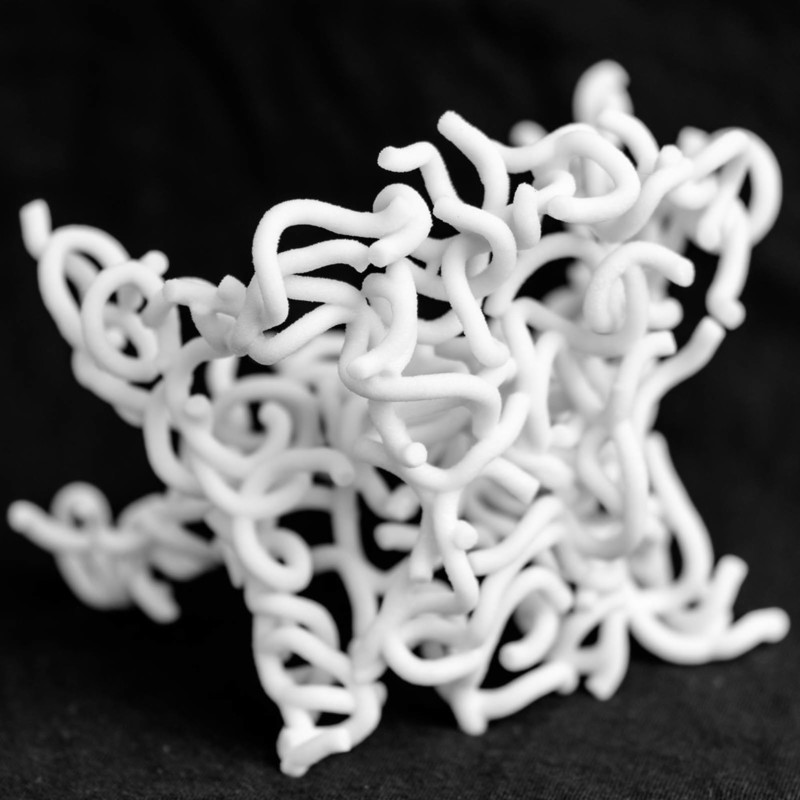}
    \includegraphics[width=0.24\linewidth]{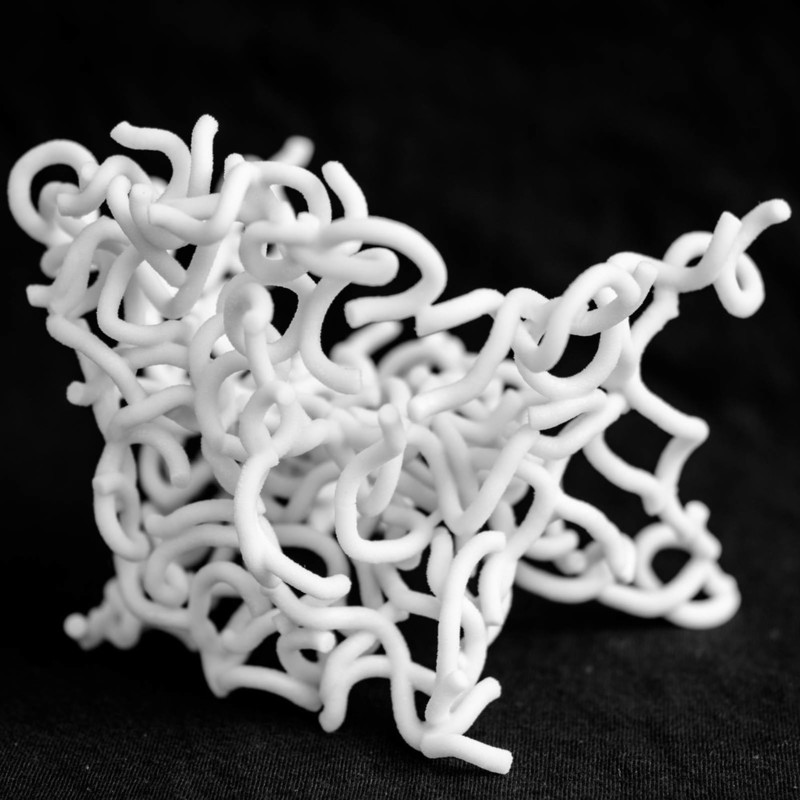}
    \caption{Different views of 3D printed chairs, the top one made of fruits printed in color, the bottom one made of deformable noodles, printed in white.}
    \label{fig:results:masterchairs}
    \label{fig:result:fruit_chair}
    \label{fig:result:masterchairs}
\end{figure*}

%% file: figs/timings.tex
\begin{table*}[htbp]
	\centering

	\providecommand{\tsec}{}
	\renewcommand{\tsec}{\,\si{\second}}
	\newcommand{\M}{\,\si{\mega}}

	\sisetup{
		table-number-alignment = center,
		table-figures-integer  = 1,
		table-figures-decimal  = 3,
		table-auto-round
	}

	\begin{tabular}
		{
		l
		S[table-format=1.2,fixed-exponent=6,table-omit-exponent,table-space-text-post={\M}]
		S[table-format=3.0]
		S[table-format=3.0]
		S[table-format=1.2]
		S[table-format=1.2]
		S[table-format=1.2]
		S[table-format=4.2]
		c
		c
		}
		\toprule
		Exemplar & {\# Vertices}& {\# Elements} & {\# Samples} & {Max (s)} & {Min (s)} & {Mean (s)} & {Total (\si{\min})} & \# Iter & Grid size \\
		\midrule
		\textbf{Chair} & & & & & & & & \\
Fruits & 2955060\M{} & 200 & 30 & 21.6176 & 5.69713 & 9.30582 & 79.0994 & 510 & $64 \times 64 \times 64$ \\
Pebbles & 1500000\M{} & 150 & 15 & 29.3928 & 5.30923 & 7.67145 & 53.7002 & 420 & $64 \times 64 \times 64$ \\
Swords & 2011520\M{} & 120 & 150 & 18.1528 & 4.5514 & 6.95507 & 48.6855 & 420 & $64 \times 64 \times 64$ \\
Wood & 13064400\M{} & 120 & 15 & 19.3875 & 5.39381 & 8.39273 & 58.7491 & 420 & $64 \times 64 \times 64$ \\
Helix & 112586\M{} & 82 & 30 & 137.581 & 16.6994 & 34.1969 & 239.378 & 420 & $96 \times 96 \times 96$ \\
		\midrule
		\textbf{Table} & & & & & & & &\\
Fruits & 2955060\M{} & 200 & 30 & 14.8394 & 3.45432 & 5.13736 & 43.6676 & 510 & $64 \times 32 \times 48$ \\
Pebbles & 2500000\M{} & 250 & 15 & 10.7681 & 3.63393 & 4.77837 & 33.4486 & 420 & $64 \times 32 \times 48$ \\
Swords & 4023040\M{} & 240 & 150 & 12.4861 & 2.8949 & 3.87392 & 27.1174 & 420 & $64 \times 32 \times 48$ \\
Wood & 17419200\M{} & 160 & 15 & 12.6272 & 3.66781 & 4.96357 & 34.745 & 420 & $64 \times 32 \times 48$ \\
		\bottomrule
	\end{tabular}

	\Caption{Timings.}{
		Number mesh vertices in the output, number of elements, number of samples per element, time per iteration (max, min and mean), total time, number of iterations and grid size.
		The number of iterations depends on the continuation parameter $\rbfWidth$ (we used $\rbfWidth_0 = 4.0$ for the fruits, $3.0$ for the rest).
	}
	\label{tab:timings}
\end{table*}

%% file: body/90-conclusion.tex
\section{Limitations and future work}
\label{sec:conclusion}

In this work we propose a novel approach for the modeling of complex aggregates of elements, which uses rigidity as a design tool, which supports rigid and deformable elements with arbitrary shapes, and synthesizes results that can be 3D printed.

%

One limitation of our approach is that there is a difference between what the optimizer sees (the underlying density grid) and the actual final geometry
(aggregate of elements). In particular, even
\begin{wrapfigure}[9]{r}{2.4cm}
	\vspace*{-4.5mm}\hspace*{-6mm}
	\includegraphics[width=3cm]{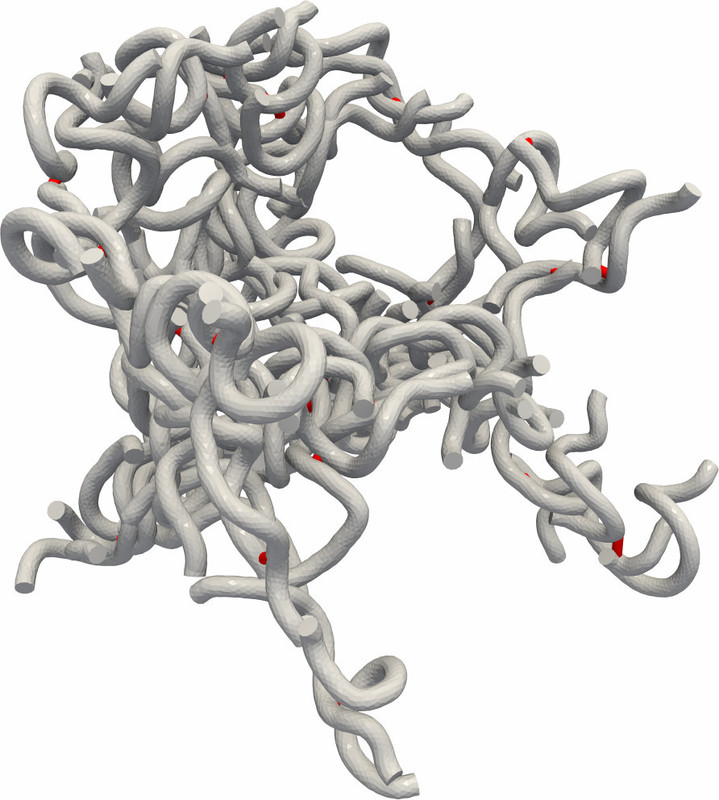}
\end{wrapfigure}
though our results typically have a single connected component there may be fragile connections between elements that barely touch each others. For this reason, it might be necessary to further reinforce the structure.

One possibility is to use the graph built during
connectivity improvement (\Cref{subsec:connectivity}) to add struts between neighboring elements. An illustrative example is shown in the inset figure (struts in red) for reinforcing a noodle chair result. Another possibility would be to slightly scale and move elements to enforce a minimal
cross-section for all contacts, or to add struts similarly to~\cite{Stava:2012:SRI} -- however a full FEM simulation would be expensive on our models.

Our method is currently an off-line synthesizer with on-line preview (each iteration takes around 10 seconds). We attempted to achieve a good balance between precision (to capture intricate geometries) and speed. However the optimization is not yet interactive.

The size of the objects we can 3D print in one piece is limited, and thus we can only produce miniatures of e.g. the chair. It would be interesting to consider printing such shapes in several parts that can be later assembled, the contacts between elements being a natural location to embed connectors.

Finally, as future work we would like to explore more user controls, possibly including pausing the optimization, making a few changes, and restarting after these additional user edits. Other controls would include direction and scale fields, as well as encouraging symmetries, to further refine the aesthetics of the results.